\documentclass{article} 
\usepackage{iclr2027_conference,times}

\usepackage{amsmath,amsfonts,bm}

\def\eqref#1{equation~\ref{#1}}

\def\1{\bm{1}}

\DeclareMathAlphabet{\mathsfit}{\encodingdefault}{\sfdefault}{m}{sl}
\SetMathAlphabet{\mathsfit}{bold}{\encodingdefault}{\sfdefault}{bx}{n}

\usepackage{graphicx}
\usepackage{hyperref}
\usepackage{url}
\usepackage{booktabs}
\usepackage{multirow}
\usepackage{makecell}
\usepackage{caption}
\usepackage{pifont}
\usepackage{siunitx}
\usepackage{graphicx}
\usepackage[table]{xcolor}
\usepackage{xcolor}

\usepackage[most]{tcolorbox}
\tcbuselibrary{listings,breakable,skins}
\newtcblisting{promptbox}[1][]{%
  breakable, enhanced, colback=gray!4, colframe=gray!55,
  boxrule=0.5pt, arc=2pt, left=4pt, right=4pt, top=3pt, bottom=3pt,
  fonttitle=\bfseries\footnotesize, coltitle=black, colbacktitle=gray!18,
  title={#1}, before skip=6pt, after skip=6pt,
  listing only,
  listing options={basicstyle=\ttfamily\scriptsize, breaklines=true,
    breakatwhitespace=false, columns=fullflexible, keepspaces=true,
    showstringspaces=false, breakindent=0pt, postbreak=\mbox{}}}

\title{OmniEcho: Audio-Visual Spatial Understanding for Omni-Modal Embodied Agents}

\author{
\rule{0pt}{2mm}\\[-5mm]
\begin{tabular}{@{}l@{}}
\bfseries
Ruixun Liu$^{1,2}$\textsuperscript{*} \quad
Yuxuan Wang$^{3}$ \quad
Jiacheng Xie$^{1}$ \quad
Yuhuan You$^{1}$\textsuperscript{*} \quad
Donghua Cai$^{4}$\textsuperscript{*} \quad
\\[1mm]
\bfseries
Junming Lin$^{4}$\textsuperscript{*} \quad
Xiong-Hui Chen$^{3}$ \quad
Zhifang Guo$^{3}$ \quad
Yunfei Chu$^{3}$ \quad
Qize Yang$^{3}$ \quad
Xize Cheng$^{3}$ \quad
\\[1mm]
\bfseries
Jin Xu$^{3}$ \quad
Yiwu Zhong$^{1,2}$\textsuperscript{\textdagger}
\\[1.5mm]
\normalfont
$^{1}$School of Intelligence Science and Technology, Peking University, \quad \\
$^{2}$State Key Laboratory of General Artificial Intelligence, Peking University, \quad \\
$^{3}$Alibaba Token Hub, Alibaba Group, \quad
$^{4}$Tsinghua University
\end{tabular}
}

\iclrfinalcopy 
\begin{document}

\maketitle

\lhead{Under review}

\renewcommand{\thefootnote}{}\footnotetext{\textsuperscript{*}Work done during internship at Alibaba Group.\hspace{1em}\textsuperscript{\textdagger}Corresponding Author.}\renewcommand{\thefootnote}{\arabic{footnote}}

\begin{abstract}
Humans can effortlessly localize the direction of a sound source and integrate it with visual cues for reasoning, yet this remains challenging for embodied agents. In particular, it is still unclear how to effectively evaluate and model spatial audio understanding in embodied settings. To address this gap, we introduce \textbf{OmniEchoBench}, a unified benchmark for spatial audio-visual perception and audio-vision-language navigation. OmniEchoBench comprises six tasks over 197 real-world spatial audio-visual scenes, 2,972 question-answer pairs, and 900 navigation samples with first-order ambisonics (FOA) audio collected from 30 real-world environments. To enable scalable training supervision, we develop a controllable rendering pipeline for spatial audio. It preserves geometric consistency among sound sources, visual observations, and agent trajectories. Building on this, we propose \textbf{OmniEcho}, a spatially aware omni-modal model. It introduces an FOA spatial encoder alongside a pretrained semantic audio pathway. 
Extensive experiments show that OmniEcho achieves state-of-the-art performance on spatial audio-visual perception. 
For our sound-guided navigation, OmniEcho reaches a performance level close to that of traditional vision-language navigation. 
These results demonstrate that spatial audio can serve as a valuable signal for embodied scene reasoning and navigation, while also highlighting fine-grained spatial localization and distance estimation as important open challenges. Our code and data will be available in \url{https://github.com/PKU-VaLuE-Lab/OmniEcho/tree/main}

\end{abstract}

\section{Introduction}
Embodied agents rely on rich multimodal signals to perceive, reason, and act. Recent advances in omni-modal understanding have substantially improved the ability to jointly process vision, language, and audio~\citep{nvidia2026nemotron3nanoomni, liu2026jaeger, qwen3omni2025, liu2025ola, ye2025omnivinci}. Meanwhile, progress in vision-language navigation (VLN) has enabled agents to reason about their surroundings and plan actions from language instructions~\citep{wei2026ground, wu2025recursive, wang2025progressthink, wei2025streamvln}. 
However, spatial audio, a common and important information source to humans, remains largely underexplored. 
Humans can effortlessly follow spatial auditory cues to localize target objects even if they are occluded, and to search for target objects beyond the current field of view. For embodied agents, such an ability is equally important. Sounds can provide information about events outside the camera's view, reveal the direction of targets, and guide agents toward their locations ~\citep{yang2024rila, chen2020soundspaces, gan2020look}. 

In this work, we address spatial audio understanding for omni-modal embodied agents. The agent receives first-order ambisonics (FOA) spatial audio together with visual observations and, when applicable, language instructions. It then reasons about the surrounding environment to answer spatial questions or navigate toward a target. 
There arise three critical challenges in this context: (1) Collecting real-world spatial-audio data is particularly costly. The received signal jointly depends on source and listener poses, scene geometry, reflections, and reverberation, which must be faithfully captured while source trajectories and listener motion are accurately annotated. (2) Simulation can provide large-scale training data. However, effective training data should not only include rendered audio, but also maintain physical consistency with visual observations and moving agents. (3) Existing pre-trained models already excel in processing omni-modal signals. How to effectively integrate spatial audio into these pre-trained models remains an open question.

\begin{figure*}[t]
    \centering
    \vspace{-2mm}
    \includegraphics[width=0.97\textwidth]{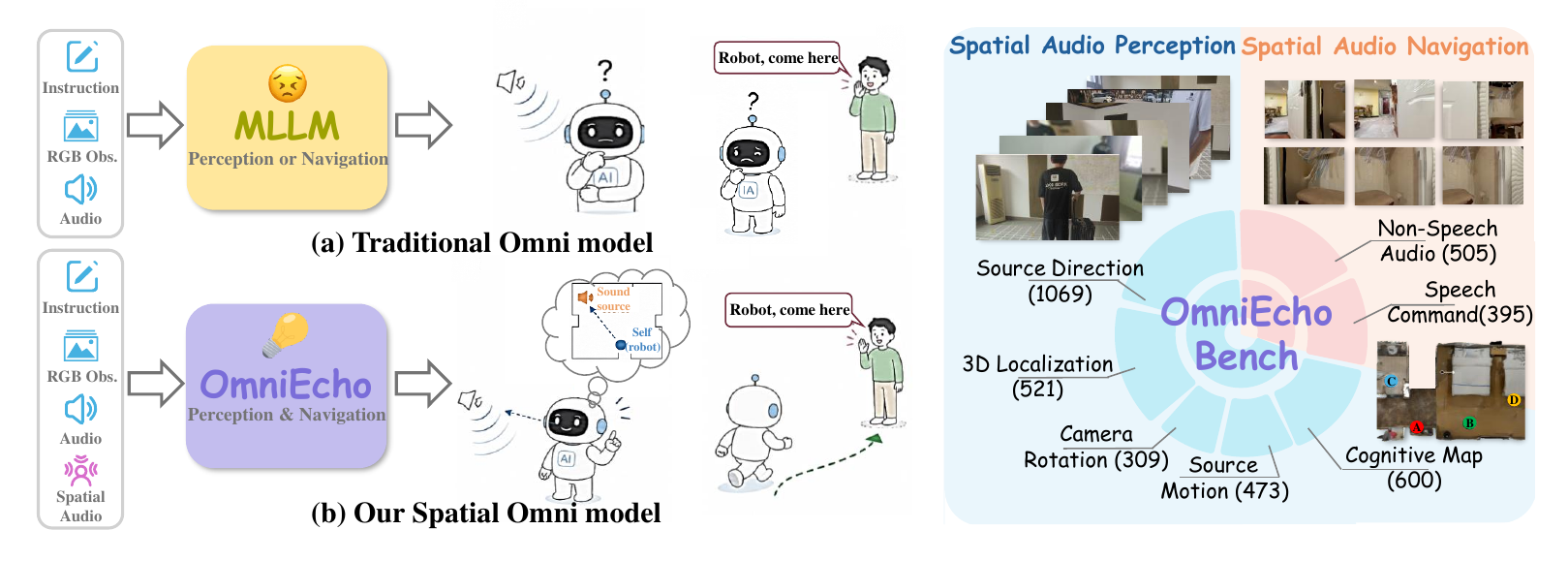}
    \vspace{-4mm}
    \caption{Overview of OmniEcho and OmniEchoBench. Left: OmniEcho extends a traditional omni model with spatial sound for perception and navigation. Right: OmniEchoBench benchmarks spatial perception and spatial audio navigation through six tasks.}
    \label{fig:overview}
    \vspace{-4mm}
\end{figure*}

To address these challenges, we first introduce \textbf{OmniEchoBench}, a unified benchmark for spatial audio-visual perception and audio-vision-language navigation, collected from real-world scenes. We also develop a scalable spatial-audio data construction pipeline. For perception, OmniEchoBench covers audio-visual scenarios involving visible, occluded, and dynamically changing sound sources, requiring agents to reason about source direction, 3D location, source motion, camera rotation, and cognitive maps~\citep{tolman1948cognitive}. For navigation, we collect spatial audio from dense positions in 30 real-world scenes and construct sound-guided navigation trajectories, together with a Habitat-based simulation platform for closed-loop evaluation~\citep{savva2019habitat}. 
Moreover, to provide scalable supervision for model development, we synthesize \textbf{large-scale training data} through controllable spatial audio rendering. For spatial perception, we construct audio-visual scenes with manually specified sound-source configurations and trajectories, and render them into FOA audio to generate spatial question-answering data. For navigation, we augment existing VLN trajectories~\citep{anderson2018vision, ku2020rxr, vlnpe, wang2023scalevln} by placing sound sources at target locations and simulating acoustic effects such as distance attenuation and room reverberation.

Building on this, we propose \textbf{OmniEcho}, a spatially aware omni-modal foundation model built on Qwen3-Omni~\citep{qwen3omni2025}. We design an FOA  encoder and introduce a three-stage training procedure to align semantic and spatial information. By preserving the model’s pretrained semantic audio pathway while introducing a spatial pathway, OmniEcho is able to jointly understand auditory semantics, spatial information, and environmental observations. 
We evaluate both spatial question answering and sound-guided navigation. Experiments show that existing omni-modal models struggle on our benchmark, whereas OmniEcho achieves state-of-the-art performance on spatial perception tasks. Although our sound-guided navigation task is more challenging than traditional text-guided VLN, OmniEcho achieves performance close to that of VLN methods. These results highlight the effectiveness of spatial audio guidance and the applicability of our task formulation.

Our main contributions are summarized as follows:
\begin{itemize}

\item To our knowledge, we are {\it the first} to address spatial audio understanding for omni‑modal embodied agents, capable of spatial audio‑visual QA and sound‑guided navigation.

\item We introduce \textbf{OmniEchoBench}, a benchmark consisting of spatial audio perception and navigation, with human annotations in real-world scenes. We further construct large-scale training data through spatial audio simulation, providing scalable training supervision.

\item We present \textbf{OmniEcho}, a foundation model that can understand spatial audio and support both spatial audio perception and navigation. Extensive experiments demonstrate its strong performance on OmniEchoBench, while revealing fundamental limitations of existing omni-modal models in spatial audio reasoning.

\end{itemize}

\section{Related Work}

\textbf{Traditional Vision-Language Navigation. } Vision-and-language navigation (VLN) requires an embodied agent to follow natural-language instructions by grounding route descriptions, landmarks, spatial relations, and action choices in visual observations and navigation history. Since Room-to-Room (R2R) introduced instruction-guided navigation in real indoor environments using the Matterport3D Simulator~\citep{anderson2018vision}, VLN has evolved from short-horizon route following toward more realistic and diagnostic navigation settings. Recent work has built fine-grained benchmarks for instruction understanding~\citep{wang2024nuances} and studied online robustness through test-time adaptation~\citep{gao2024fastslow} and obstructed environments~\citep{hong2024obstructed}. Other studies have improved reasoning or memory through causal learning, adaptive history filtering, language-based perceptual representations, explicit LLM reasoning, and bidirectional cross-modal modeling~\citep{wang2024causal,he2024memory,pan2024langnav,zhou2023navgpt,wu2025recursive}. Despite these advances, representative VLN models remain fundamentally vision-text driven: their decisions mainly depend on RGB/RGB-D observations, textual instructions, object or landmark descriptions, navigation histories, or language-converted scene representations. They do not treat spatial acoustic cues, such as sound-source direction, distance, and spatial layout, as native evidence for navigation. We instead introduce a navigation model with spatial audio understanding capability.

\textbf{Audio-Visual Navigation. } Audio-visual navigation studies embodied agents that use auditory and visual observations to reach sound-emitting targets in 3D environments. Look, Listen, and Act first formalized audio-visual embodied navigation as shortest-path planning toward a sound source from egocentric audio-visual inputs~\citep{gan2020look}. SoundSpaces and SoundSpaces 2.0 then established widely used benchmarks by adding geometry-based acoustic rendering to Habitat over Matterport3D and Replica. However, these benchmarks rely on rendered acoustics and do not cover spoken-instruction navigation~\citep{chen2020soundspaces,chen2022soundspaces}. Subsequent work improved waypoint selection, acoustic memory, semantic sound handling, and robustness to moving or noisy sources~\citep{chen2021waypoints,chen2021semantic,younes2023catch,chen2023oran,shi2025enmus}. Other studies explored language-mediated interaction or planning for audio-visual navigation, including AVLEN, CAVEN, AVLMaps, and RILA~\citep{paul2022avlen,liu2024caven,huang2023avlmaps,yang2024rila}. However, these studies do not  evaluate navigation in scenarios with real-world recorded spatial audio. In contrast, we introduce a unified benchmark for simultaneously evaluating spatial audio perception and navigation, and we provide a navigation simulator built upon real-world collected  FOA audio, thereby offering more realistic and reliable acoustic evidence for navigation.

\textbf{Spatial Audio Understanding. } 
Recent work has improved spatial acoustic modeling from three complementary angles. One line develops encoders that use binaural phase differences, intensity cues, GCC-PHAT features, microphone geometry, or task-disentangled branches, improving localization, distance estimation, and spatial reasoning~\citep{zheng2024bat,wilkinghoff2025dspast,dementyev2026phasecoder}. A second line aligns spatial audio with language, either through contrastive learning over open-vocabulary text or through structured embeddings that separate semantic and spatial factors for retrieval, understanding, and editing~\citep{devnani2024elsa,hu2025salm}. A third line connects spatial audio representations to audio-language models for natural-language description and reasoning about direction, distance, reverberation, and multi-source relations~\citep{sakshi2025spur,jiang2025sciphi,biswas2026owl,you2026world,liu2026jaeger}. Overall, these works demonstrate that spatial audio can be encoded, aligned with language, and used for language-based spatial reasoning. Recent work has begun to equip MLLMs with spatial audio understanding. The closest related work is Spatial-Omni~\citep{zhu2026spatialomni}, which explores spatial audio understanding. However, it does not incorporate visual frames during training and does not address sound-guided navigation. In contrast, our benchmark uses real-world FOA recordings and evaluates both spatial audio-visual understanding and sound-guided navigation.

\section{Benchmark and Datasets}
\textbf{Overview.} This section presents the benchmark and dataset constructed in this work, including OmniEchoBench-QA for evaluating spatial-audio understanding and OmniEchoBench-Nav for assessing sound-guided navigation. It further describes the training data synthesis pipeline, which is built upon a unified FOA spatial-audio rendering framework to generate spatial audiovisual question-answering data, audio-only data, and sound-guided navigation data.
\subsection{OmniEchoBench-QA}
\begin{figure*}[t]
    \centering
    \vspace{-4mm}
    \includegraphics[width=0.95\textwidth]{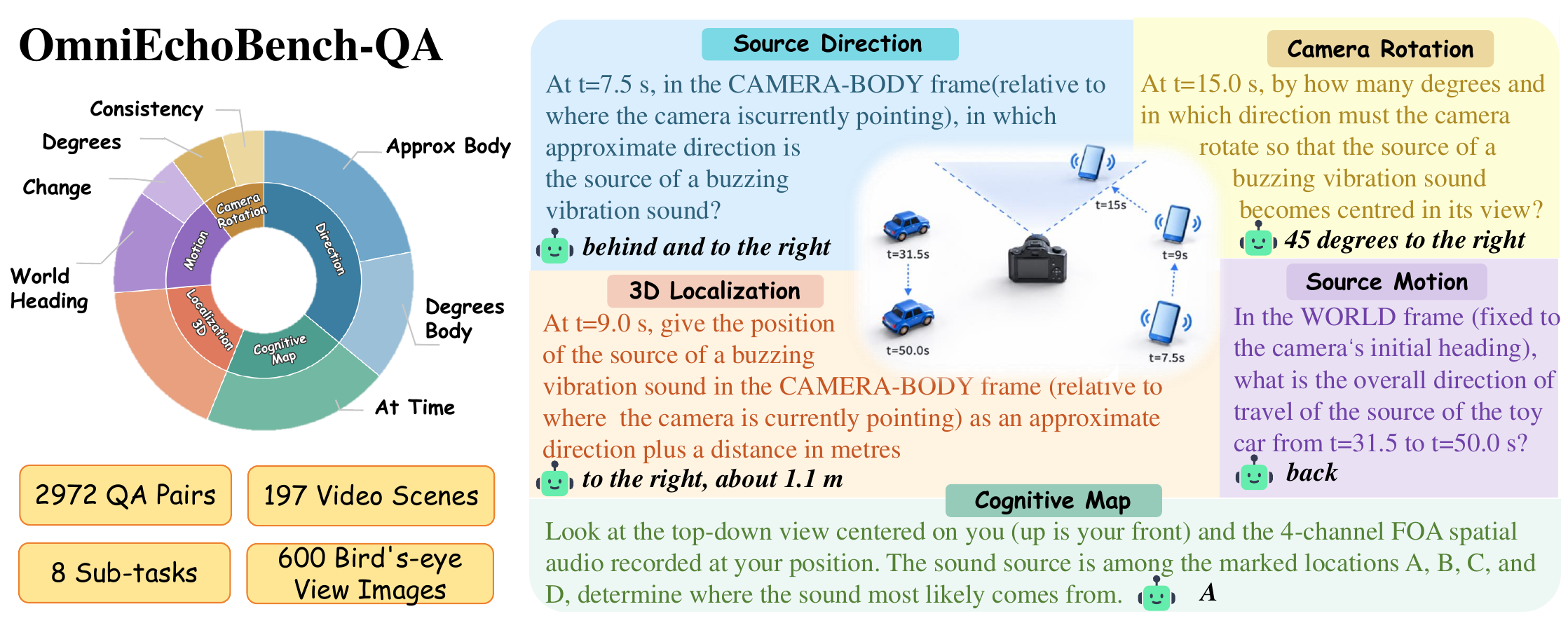}
    \vspace{-3mm}
    \caption{{Overview of OmniEchoBench-QA, which comprises 2,972 QA pairs across 197 scenes and 600 bird's-eye view images, assessing a model’s understanding of spatial audio.}}
    \vspace{-4mm}
    \label{fig:benchmark-QA}
\end{figure*}

\begin{figure*}[t]
    \centering
    \includegraphics[width=0.95\textwidth]{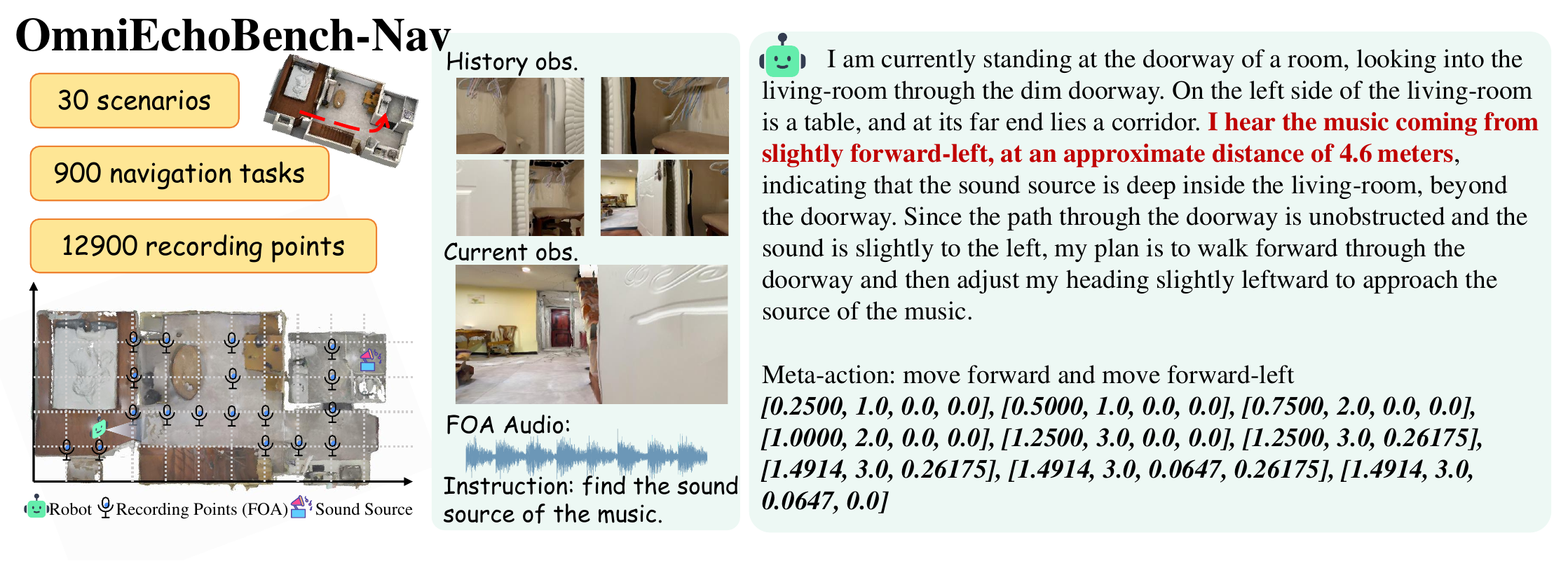}
    \vspace{-4mm}
    \caption{Overview of OmniEchoBench-Nav. Collected across 30 real-world scenes, the benchmark contains 900 navigation tasks, where agents use historical frames, the current view, FOA audio, and an instruction to navigate toward the target.}
    \vspace{-5mm}
    \label{fig:benchmark-Nav}
\end{figure*}
\textbf{Design principle.}
As shown in Figure~\ref{fig:benchmark-QA}, we introduce OmniEchoBench-QA to evaluate whether an omni-modal model can perceive and reason about sound-source geometry from egocentric observations. This benchmark emphasizes sound sources that lie outside the current visual field or dynamically enter and leave the field of view over time. Unlike synthetic setups, all data are collected from real human performances. We first author a set of scripts (\emph{screenplays}) that specify the acoustic events, the actors' motion, and their spatio-temporal relation to the camera (see the accompanying screenplay sheet), deliberately emphasizing off-screen and boundary-crossing sound sources so that the visual stream alone is insufficient for answering the questions. Human actors then perform and record each script, capturing, in a single synchronized take, a first-person main-view video, a $360^{\circ}$ panoramic video, and a four-channel first-order Ambisonics (FOA, ACN/SN3D) audio track. The panoramic video is used \emph{only} during annotation; at test time, the model is given a single-view video (or a still image) together with the FOA audio, matching a realistic embodied-perception setting.

\textbf{Annotation and question construction.}
Using the panoramic recording as the ground-truth reference, annotators label the real-time position of every active sound source at $2\,\mathrm{Hz}$, yielding a dense spatio-temporal trajectory for each source relative to the recording device. Then we apply rule-based sampling to instantiate questions in several typed categories, automatically deriving the ground-truth answers. The benchmark comprises two complementary subsets and $2{,}972$ QA pairs in total. The first subset ($2{,}372$ QA) probes fine-grained temporal spatial reasoning over the recorded videos: \emph{source direction} ($1{,}069$), \emph{3D localization} ($521$), \emph{source motion} ($473$), and \emph{camera rotation} ($309$). 
The second subset ($600$ QA) targets audio-driven source localization from a bird's-eye view. In this setting, the model is presented with an ego-centered top-down map (with ``up'' aligned to its forward direction) together with the co-located FOA clip collected in navigation scenes. The model must then select the most plausible sound-source location from four candidates marked ${A, B, C, D}$. This setup more directly isolates the model's ability to integrate spatial hearing with a cognitive map. 
Additional information can be found in the Appendix~\ref{app:bench-qa}.

\subsection{OmniEchoBench-Nav}

\textbf{Design and Annotation.} As shown in Figure~\ref{fig:benchmark-Nav}, we introduce OmniEchoBench-Nav, a real-world indoor navigation benchmark that jointly supports \emph{spatial-audio} and \emph{language-instruction} navigation over the same set of scenes and targets. The benchmark comprises $30$ real-scanned indoor environments, distributed over single-room ($10$) and multi-room cross-room layouts ($20$, spanning two to four connected rooms), each provided as a textured mesh (\texttt{.glb}) together with a top-down floor render.  Each scene contains $30$ navigation samples, giving $900$ samples in total. In each sample, a single sound source is placed at a known $3$D position. Its signal is captured at a dense grid of receiver positions with known coordinates, producing $12{,}900$  FOA recordings ($4$-channel, $48$\,kHz) that together form a spatially dense acoustic field of the environment. Sources cover a broad range of everyday audio, split into non-speech environmental events (e.g., mechanical, alarm, doorbell, animal, footsteps; $505$ samples) and spoken commands (e.g., help requests, calls; $395$ samples). Each source is annotated with its source device, room, semantic description, and a natural-language description of the navigation target. At test time, the agent is provided with the spatial audio recorded at the receiver position nearest to its current location, with directional remapping performed according to its current orientation. Based on this perceived spatial audio, the embodied agent must infer the location of the target sound source and navigate toward it. The Appendix~\ref{app:bench-nav} describes the real-world scenes and acoustic-field collection process in greater detail.

\label{sec:pipeline}

\begin{figure*}[t]
    \centering
    
    \includegraphics[width=\textwidth]{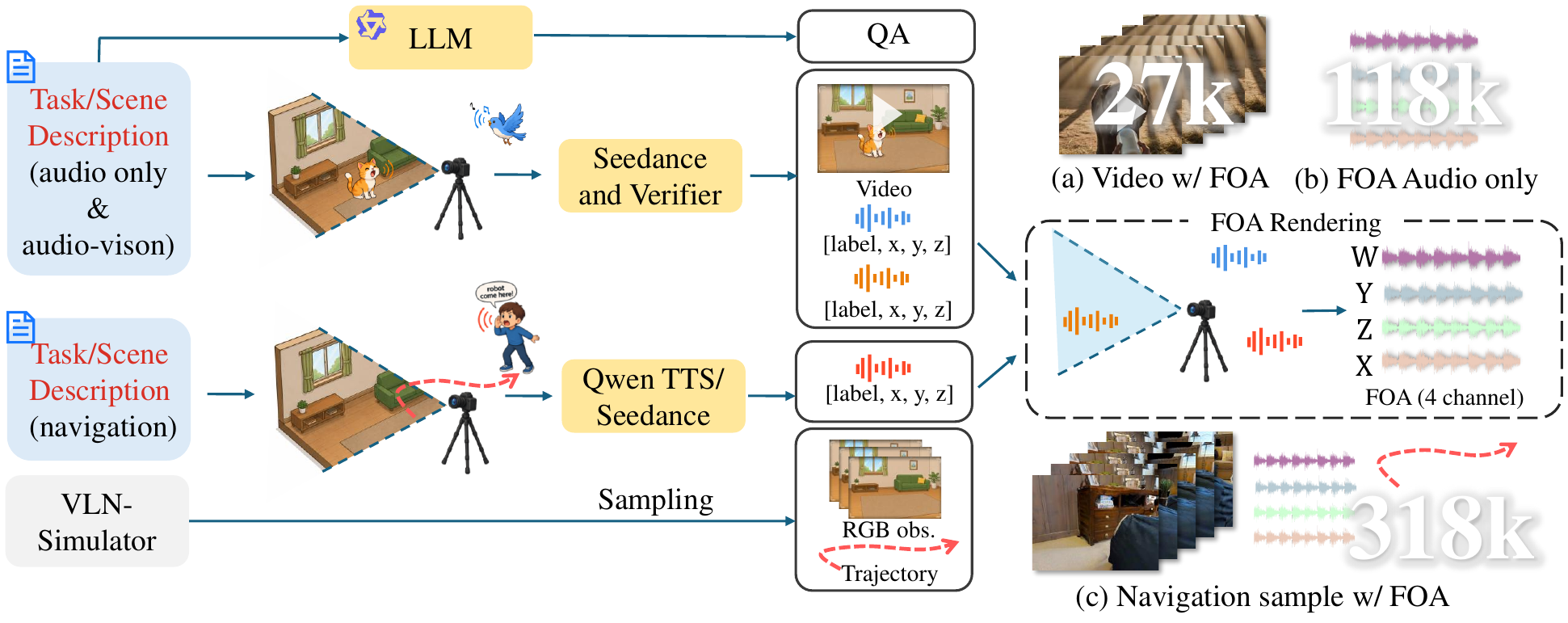}
    \vspace{-8mm}
    \caption{Overview of our training data generation pipeline. From simulated scenes, the system generates QA pairs, videos and sound event annotations and renders mono audio into FOA data.}
    \vspace{-3mm}
    \label{fig:synthesis-pipeline}
\end{figure*}
\subsection{Spatial Audiovisual Simulation for Data Synthesis}

We build all spatial perception and navigation data on top of a unified spatial-audio rendering framework. Each sounding source is represented by a time-varying 3D trajectory, while the listener is co-located with the camera. We adopt a right-handed listener frame with the camera at the origin facing \(+y\); azimuth \(\theta \in [-180^\circ,180^\circ]\) is defined with \(0^\circ\) straight ahead, \(+90^\circ\) to the left, and \(-90^\circ\) to the right. Given the source trajectories and listener pose, we encode each monophonic source stem into FOA (4 channels, ACN/SN3D) using real spherical harmonics~\citep{nachbar2011ambix,zotter2019ambisonics}:
\begin{equation}
  \mathbf{a}(t) = \sum_i \frac{g_i}{r_i(t)}
  \mathbf{Y}\!\big(\theta_i(t),\phi_i(t)\big)s_i(t),
  \label{eq:foa}
\end{equation}
where \(s_i(t)\), \(r_i(t)\), and \(g_i\) denote the source stem, source--listener distance, and source gain, respectively. Source directions are transformed into the current listener frame to maintain spatial alignment with the visual observations, with optional distance attenuation and room effects. This unified formulation is shared by both spatial audiovisual perception and sound-guided navigation.

Figure~\ref{fig:synthesis-pipeline} illustrates two training-data pipelines. For spatial audiovisual QA, we generate dynamic scenes with Seedance~\citep{seedance2026seedance} and combine their visual content with trajectory-grounded sound events, from which FOA audio and QA pairs are synthesized. We further synthesize audio-only training data to support FOA encoder pretraining and QA. For navigation, we augment existing VLN trajectories with destination-conditioned sound events and render the corresponding FOA observations along sampled trajectories. The datasets provide synchronized visual observations, spatial audio, and geometry-grounded supervision for both perception and navigation. Further details of the data generation and annotation procedures are provided in the Appendix~\ref{app:data-generation}.

\section{Method}
\subsection{Task Formulation}
\begin{figure*}[t]
    \vspace{-2mm}
    \centering
    \includegraphics[width=0.92\textwidth]{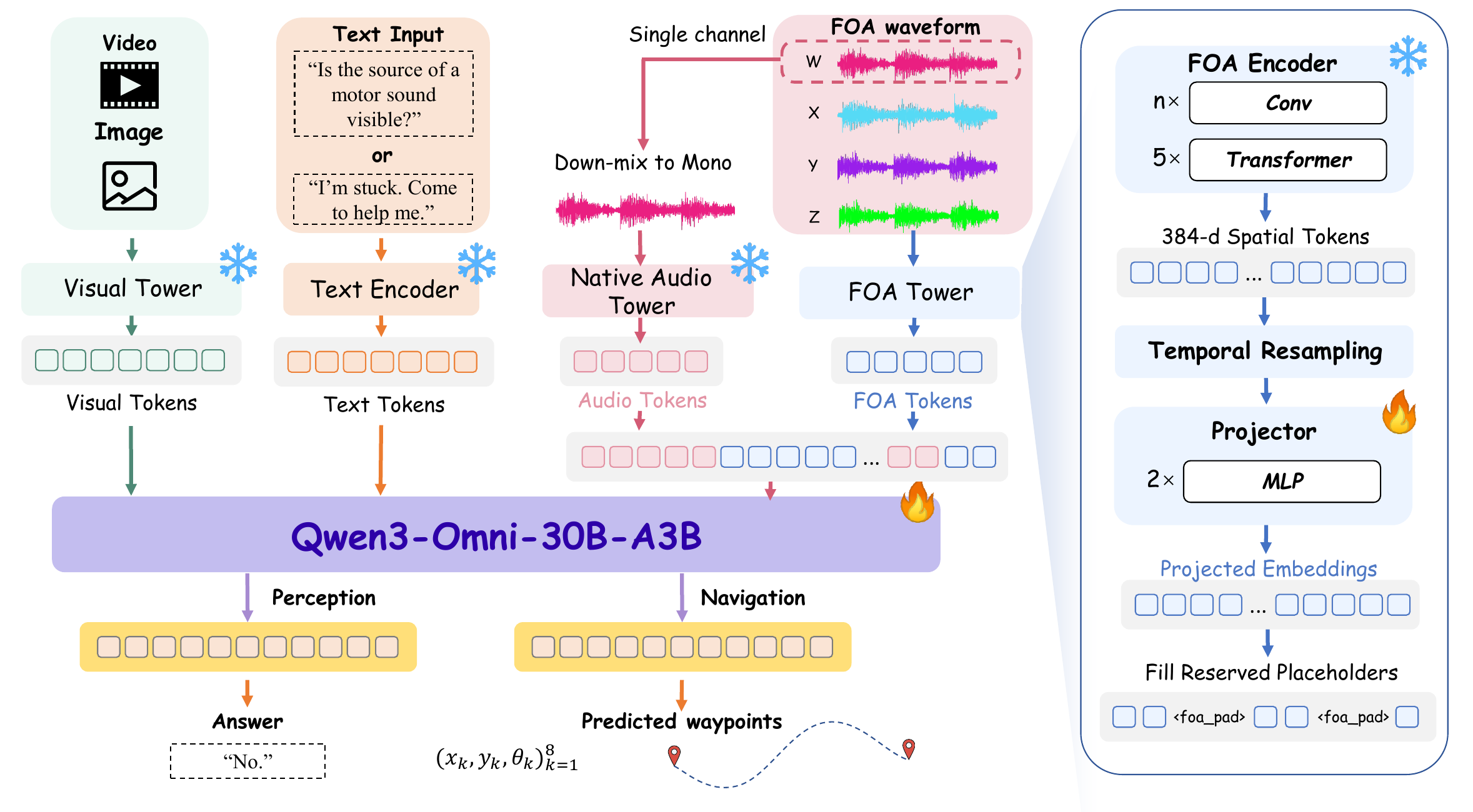}
    \vspace{-5mm}
    \caption{Overview of our unified spatial omni model built on Qwen3-Omni-Thinker. The pretrained FOA encoder extracts spatial tokens; these tokens are  projected into the language embedding space by a trainable projector, and inserted after the corresponding semantic audio tokens. }
    \vspace{-5mm}
    \label{fig:model-architecture}
\end{figure*}
We unify spatial audio-visual perception and navigation within a single embodied-agent framework. For perception tasks, the model takes FOA  audio \(A\), a video sequence \(V\) or a single image \(I\), and a question \(q\), and predicts an answer \(a\).
For navigation tasks, we follow the setting of InternNav~\citep{wei2026ground}. At each decision step, the model receives a language instruction \(L\), \(T\) frames captured by a forward-facing camera, and the FOA spatial audio \(A_t\) within the current temporal window, and predicts a trajectory over the next \(K=8\) time steps. 
We uniformly sample \(T-1\) keyframes from past observations and combine them with the current frame as visual input.
This formulation enables unified spatial audio-visual perception and navigation within a single framework.

\subsection{Spatial audio encoding}
\textbf{Stage 1: Semantic alignment of the FOA encoder.}
We first pretrain a lightweight FOA encoder $f_\theta$ that maps an FOA clip into a temporal sequence of $d$-dimensional tokens ($d=384$). Each clip is converted into a $5$-channel input map $\mathbf{X}\in\mathbb{R}^{5\times128\times T}$ comprising the log-mel spectrogram of the omnidirectional ($W$) channel together with the three active-intensity components $(i_x,i_y,i_z)$ and the diffuseness $\delta$, all in the ACN/SN3D convention. The encoder is trained to align its representations with a frozen CLIP text encoder~\citep{radford2021clip} using a SigLIP objective~\citep{zhai2023siglip}, thereby establishing an open-vocabulary sound-semantic space. Training uses a 100k-clip corpus of synthetic FOA scenes containing 1--4 sound sources with diverse spatial configurations.

\textbf{Stage 2: Query-conditioned spatial localization.}
Starting from the semantically aligned encoder, we further fine-tune it for query-conditioned sound-source localization. Given a text query specifying a target source, a cross-attention localization head attends to the encoder tokens and predicts the source's azimuth, elevation, and distance. The model is optimized with a localization loss together with the Stage-1 semantic objective to preserve the learned sound semantics.

\textbf{Stage 3: Integration into the Omni backbone.}
As shown in Figure~\ref{fig:model-architecture}, we graft the frozen FOA encoder $f_{\theta^\star}$ into the Qwen3-Omni model~\citep{qwen3omni2025}. Each FOA clip is processed along two complementary paths: (i) its $W$ channel is down-mixed to mono and fed to the original, \emph{frozen} audio tower to preserve the model's native semantic audio tokens; and (ii) the same clip is passed through $f_{\theta^\star}$ to obtain spatial tokens, which are temporally resampled to the $7$\,Hz audio-token grid and mapped into the language embedding space by a trainable projector. The projected spatial tokens occupy reserved placeholder embedding rows (\texttt{<foa\_pad>}) and are inserted immediately after the corresponding audio tokens in temporal order. Only the LLM parameters $\Psi$ and the projector $\phi$ are updated, while the FOA encoder, audio tower, and visual tower remain frozen. 
Additional encoder architecture and optimization details for all three training stages are included in the Appendix~\ref{app:training-details}.

\begin{table}[t]
    \centering
    
    \fontsize{6.3pt}{8.4pt}\selectfont
    \setlength{\tabcolsep}{3.5pt}
    \renewcommand{\arraystretch}{1.15}
    
    \caption{\textbf{Results on our OmniEchoBench-QA}. Our OmniEcho performs strongly in both audio-only and audio-visual settings, outperforming most evaluated models in overall accuracy and across the individual spatial reasoning tasks. 'A-single' denotes single-channel audio, 'A-foa' denotes four-channel FOA audio, and 'V' denotes visual input.  }
    \vspace{-4mm}
    \label{tab:main1}
    \scalebox{0.95}{
    \begin{tabular}{l l l c c c c c c}
        \toprule
        \makecell[c]{\textbf{Method}} &
        \makecell[c]{\textbf{Base Model}} &
        \makecell[c]{\textbf{Setting}} &
        \makecell[c]{\textbf{Camera}\\\textbf{Rotation}} &
        \makecell[c]{\textbf{Source} \\ \textbf{Direction}} &
        \makecell[c]{\textbf{3D}\\\textbf{Localization}} &
        \makecell[c]{\textbf{Source}\\ \textbf{Motion}} &
        \makecell[c]{\textbf{Cognitive}\\\textbf{Map}} &
        \makecell[c]{\textbf{Overall}} \\
        \midrule

        Random & - & - & {33.6} & 12.5 & - & {23.8}  & {25.0} & - \\
        \midrule
        \rowcolor{gray!15}
        \multicolumn{9}{c}{\textbf{Audio only}} \\
        Qwen3-Omni-30B-A3B~\citep{qwen3omni2025} & Qwen3-Omni-30B-A3B & A-single & 22.7 & 18.4 & 8.3 & 22.0 & - & 17.5 \\
        Qwen2.5-Omni-7B~\citep{qwen25omni2025} & Qwen2.5-Omni-7B & A-single & 22.7 & 8.3 & 4.6 & 14.6 & & 10.6 \\
        GLM4-Voice-9B~\citep{zeng2024glm4voice} & GLM4-9B & A-single & 20.4 & 9.5 & 5.4 & 20.3 & - & 12.2 \\
        SO-7B~\citep{zhu2026spatialomni} & Qwen2.5-Omni-7B & A-foa & 21.4 & 7.1 & 0.0 & 20.7 & - & 10.1 \\
        
        OmniEcho (Ours) & Qwen3-Omni-30B-A3B & A-foa  & \textbf{29.1} & \textbf{22.1} & \textbf{14.0} & \textbf{24.9} & - & \textbf{21.8}  \\
        \midrule
        \rowcolor{gray!15}
        \multicolumn{9}{c}{\textbf{Audio \& Vision}} \\

        Qwen3-Omni-30B-A3B~\citep{qwen3omni2025} & Qwen3-Omni-30B-A3B & A-single + V & {32.7} & {16.3} & {7.7} & {20.9}  & {22.3}  & {18.5} \\
        Qwen2.5-Omni-7B~\citep{qwen25omni2025} & Qwen2.5-Omni-7B & A-single + V & 24.9 & 8.0 & 5.4 & 14.0  & 20.0 & 12.6 \\

        SO-7B~\citep{zhu2026spatialomni} & Qwen2.5-Omni-7B & A-foa + V & 21.4 & 5.2 & 0.0 &  17.1 & 22.8 & 11.4 \\
        
        
        OmniEcho (Ours) & Qwen3-Omni-30B-A3B & A-foa + V  & \textbf{41.7} & \textbf{22.6} & \textbf{14.2} & \textbf{24.7} & \textbf{47.5} & \textbf{28.5}  \\

        \bottomrule
    \end{tabular}
    }
    \vspace{-2mm}
    
\end{table}

\begin{table}[t]
    \centering
    \fontsize{7pt}{8.4pt}\selectfont
    \setlength{\tabcolsep}{3.5pt}
    \renewcommand{\arraystretch}{1.15}
    \caption{\textbf{Results on our OmniEchoBench-Nav.} Our benchmark is designed to evaluate sound-guided navigation, and meanwhile, additionally provides text instruction for traditional text-guided VLN methods. Even if audio guidance is more challenging than text instruction, our method achieves close performance as previous VLN methods.}

    \vspace{-4mm}
    \label{tab:main2}

    \begin{tabular}{@{}l c c c c c c c@{}}
        \toprule
        \multirow{2}{*}{\textbf{Model}}
        & \multirow{2}{*}{\textbf{Base Model}}
        & \multirow{2}{*}{\textbf{Guidance}}
        & \multicolumn{5}{c}{\textbf{Val-Unseen}}
        \\
        \cmidrule(lr){4-8}

        & & & \textbf{SR$\uparrow$}
        & \textbf{SPL$\uparrow$}
        & \textbf{NE$\downarrow$}
        & \textbf{OS$\uparrow$}
        & \textbf{TL}
        \\
        \midrule

        Seq2Seq~\citep{krantz2020vlnce}
        & End-to-end
        & Text Instruction
        & 11.3
        & 9.4
        & {4.14}
        & {53.00}
        & 5.51
        \\

        CMA~\citep{krantz2020vlnce}
        & End-to-end
        & Text Instruction
        & 9.8
        & 8.5
        & 4.34
        & 48.56
        & 5.00
        \\

        InternVLA-N1~\citep{internrobotics2025internvla}
        & Qwen2.5-VL-7B
        & Text Instruction
        & {17.8}
        & {16.7}
        & {2.87}
        & {63.89}
        & 4.67
        \\

        \midrule
        Soundspaces~\citep{chen2020soundspaces}
        & End-to-end
        & Binaural  audio
        & 5.4
        & 3.9
        & 5.32
        & 25.67
        & 4.16
        \\

        OmniEcho
        & Qwen3-Omni-30B-A3B
        & FOA audio
        & {16.2}
        & {11.5}
        & 4.09
        & 52.89
        & 7.02
        \\

        \bottomrule
    \end{tabular}
    \vspace{-6mm}
\end{table}

\begin{table}[t]
    \centering
    \fontsize{6.8pt}{8.4pt}\selectfont
    \setlength{\tabcolsep}{3.5pt}
    \renewcommand{\arraystretch}{1.15}
    \caption{\textbf{Performance on VLN-CE R2R with simulated spatial audio.} Although sound-guided navigation is more challenging than text-guided VLN, our method achieves comparable performance to previous baselines, demonstrating the potential of our model and dataset for future research.}
    \label{tab:main3}
    \vspace{-4mm}
    \scalebox{0.95}{
    \begin{tabular}{@{}l c c c c c c c@{}}
        \toprule
        \multirow{2}{*}{\textbf{Model}}
        & \multirow{2}{*}{\textbf{Base Model}}
        & \multirow{2}{*}{\textbf{Guidance}}
        & \multicolumn{5}{c}{\textbf{Val-Unseen}}
        \\
        \cmidrule(lr){4-8}

        & & & \textbf{SR$\uparrow$}
        & \textbf{SPL$\uparrow$}
        & \textbf{NE$\downarrow$}
        & \textbf{OS$\uparrow$}
        & \textbf{TL}
        \\
        \midrule

        Seq2Seq~\citep{krantz2020vlnce}
        & End-to-end
        & Text Instruction
        & 21.0
        & -
        & 7.81
        & 28.0
        & 8.39
        \\

        CMA~\citep{krantz2020vlnce}
        & End-to-end
        & Text Instruction
        & 22.0
        & 20.0
        & 8.17
        & 28.0
        & 7.42
        \\

        VLN-R1~\citep{qi2025vlnr1}
        & Qwen2-VL-7B~\citep{qwen2vl2024}
        & Text Instruction
        & 30.2
        & 21.8
        & 7.0
        & 41.2
        & 10.0
        \\

        InternVLA-N1~\citep{internrobotics2025internvla}
        & Qwen2.5-VL-7B~\citep{qwen25vl2025}
        & Text Instruction
        & {50.6}
        & {43.3}
        & {4.73}
        & {56.7}
        & -
        \\

        NAViLA-SAGE~\citep{miao2026towards}
        & navila-siglip-llama3-8b-v1.5-pretrain
        & Text Instruction
        & {38.0}
        & {36.0}
        & -
        & {51.0}
        & -
        \\
        \midrule
        OmniEcho
        & Qwen3-Omni-30B-A3B~\citep{qwen3omni2025}
        & Audio Instruction
        & 22.2
        & 16.3
        & 7.31
        & 24.5
        & 5.54
        \\

        \bottomrule
    \end{tabular}}
    \vspace{-6mm}
\end{table}

\section{Experiments}
\subsection{Setup}

\textbf{Implementation Details.} We build our model upon the pretrained Qwen3-Omni-30B-A3B~\citep{qwen3omni2025}. Specifically, the model is trained on spatial audio-visual QA together with navigation-oriented supervision, resulting in 363,193 training examples in total. The navigation subset consists of 139,142 examples from ScaleVLN~\citep{wang2023scalevln}, 92,418 from RxR~\mbox{\citep{ku2020rxr}}, 31,577 from R2R~\citep{anderson2018vision}, and 28,969 from VLN-PE~\citep{vlnpe}.
Stage~3 training is conducted on 32 NVIDIA A100 GPUs and requires approximately four days. 

\textbf{Benchmarks.}  For spatial audio-visual perception, we report results on OmniEchoBench-QA. For navigation, we evaluate on OmniEchoBench-Nav as well as the VLN-CE R2R val unseen split \citep{anderson2018vision,krantz2020vlnce}. More details are documented in the Appendix~\ref{app:evaluation-details}.

\subsection{Main results}

\subsubsection{Spatial Audio-Visual Question Answering Results}

Table~\ref{tab:main1} reports the results on OmniEchoBench-QA. Under both the audio-only and audio-visual settings, OmniEcho outperforms evaluated models overall and across individual tasks. In the audio-only setting, OmniEcho surpasses the Qwen3-Omni-30B-A3B~\citep{qwen3omni2025} backbone which has only  single-channel audio encoder, demonstrating the effectiveness of our FOA encoder. OmniEcho significantly outperforms SO-7B~\citep{zhu2026spatialomni} on our real-world collected benchmark. Although SO-7B claims to employ spatial-audio training, it lacks visual supervision during training and still suffers from a sim-to-real gap. These findings suggest that existing methods remain limited in their robustness for spatial-audio understanding in real-world scenarios.

Comparing OmniEcho across the two input settings, removing visual input leaves its performance on Source Direction, 3D Localization, and Source Motion largely unchanged, indicating that the model can perform these forms of spatial reasoning primarily from FOA audio. Camera Rotation, however, degrades substantially without vision, suggesting that this task depends more strongly on visual evidence of viewpoint changes. This pattern is consistent with the FOA-encoder ablation, where explicit FOA features benefit source-centric spatial tasks but provide limited gains for camera-rotation reasoning. Together, these results suggest that FOA primarily captures sound-source geometry, while visual observations remain important for camera-centric reasoning.

We acknowledge that training with synthetically generated simulation data introduces bias. We align the distribution of simulated sound-source states with that of real-world states. Empirically, the simulated data has been shown to endow the model with a fundamental capability for spatial audio perception. In addition, the experiments in Appendix~\ref{app:sim2real} suggest that, in the future, the training distribution could potentially be corrected using a small amount of real spatial-audio data.

\subsubsection{Spatial Audio Navigation Results}

\textbf{Baseline Selection.} SoundSpaces-based models, AVLEN, and RILA~\citep{chen2020soundspaces,paul2022avlen,yang2024rila} are originally designed for simulated acoustic observations and do not support FOA input. To provide an informative comparison, we include the open-source SoundSpaces model and convert the FOA audio into binaural audio. Nevertheless, some navigation instructions in OmniEchoBench-Nav are conveyed through the spatial audio itself, requiring the agent not only to pursue a sound-emitting target but also to perform joint semantic understanding and spatial grounding. Therefore, the task formulations of these methods cannot be fully aligned with ours.

\textbf{Results on OmniEchoBench}. Table~\ref{tab:main2} reports the results on the OmniEchoBench-Nav. Following text-guided models, we adopt the agent-stopping setting, in which each model autonomously decides when to terminate. OmniEcho achieves 16.2\% SR and 11.5\% SPL, substantially outperforming the conventional text-guided baselines Seq2Seq and CMA. Compared with InternVLA-N1~\citep{internrobotics2025internvla}, OmniEcho is lower by 1.6 points in SR and 5.2 points in SPL. These results indicate that, without route-level textual guidance or goal images, sound-guided navigation in real-world environments can still approach the performance of some traditional text-guided baselines. In contrast, SoundSpaces with binaural audio performs markedly worse in real spatial-audio settings and struggles to determine when to stop autonomously. For OmniEcho and other spatial-audio navigation models, efficient navigation and reliable stopping remain significant challenges.

\textbf{Results on VLN-CE R2R}. Table~\ref{tab:main3} further reports results on the simulated VLN-CE R2R Val-Unseen split~\citep{krantz2020vlnce}. We again use agent stopping for the cross-model comparison. Our OmniEcho achieves an SR of 22.2\%, slightly exceeding Seq2Seq and CMA~\citep{krantz2020vlnce}, although its SPL of 16.3\% is below CMA (20.0\%). Its NE of 7.31 is better than those of Seq2Seq and CMA, and its trajectory length of 5.54 is the shortest among methods with reported TL. Nevertheless, OmniEcho remains behind VLN-R1~\citep{qi2025vlnr1}, NAViLA-SAGE~\citep{miao2026towards}, and InternVLA-N1~\citep{internrobotics2025internvla} in SR and SPL. Overall, spatial-audio guidance is competitive with earlier text-guided VLN baselines in success rate, but a clear gap remains relative to stronger text-guided systems, especially in path efficiency. We provide additional analyses of navigation failure modes and the effect of ground-truth sound direction in Appendix~\ref{app:vln-failure-modes} and~\ref{app:vln-gt-direction}.

\subsection{Ablation study}

\textbf{Ablation on native audio encoder and FOA encoder.} Table~\ref{tab:ablation} ablates three designs in Qwen3-Omni-30B-A3B~\citep{qwen3omni2025}: introducing the FOA encoder, retaining the native audio encoder, and freezing the FOA encoder during Stage~3. The full configuration achieves the best Overall score of 28.5. Adding the frozen FOA encoder to the native-audio-only baseline improves Overall and most subtasks, confirming that it provides essential spatial cues beyond those captured by the native audio pathway. The slight decrease on Camera Rotation suggests that extra visual observations reduce reliance on spatial audio cues. Removing the native audio encoder while retaining the frozen FOA encoder reduces Overall from 28.5 to 26.2, showing that native semantic audio tokens remain complementary to the spatial representation. Finally, updating rather than freezing the FOA encoder during Stage~3 degrades every subtask, suggesting that freezing better preserves the spatial representations from the first two stages and avoids interference during multimodal instruction tuning.

\begin{figure*}[t]
    \centering
    \includegraphics[width=0.9\textwidth]{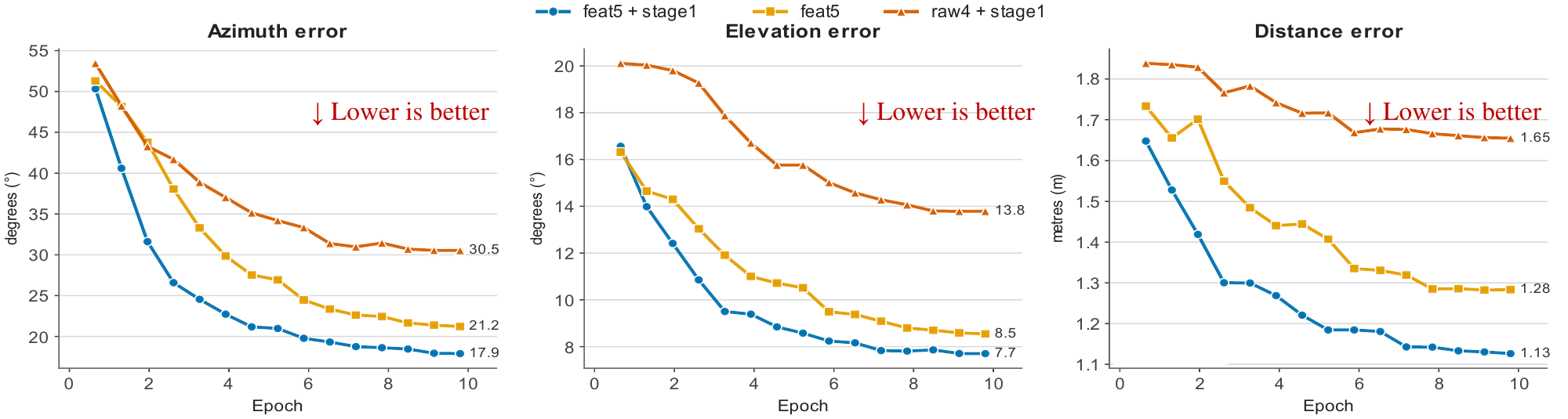}
    \vspace{-3mm}
    \caption{{\textbf{Ablation on input representation and Stage-1 pretraining.}  feat5+stage1 achieves the best azimuth, elevation, and distance errors, outperforming raw4+stage1 and feat5 (w/o stage 1 pretraining), showing that feature-based input is more effective than raw FOA waveform input and that Stage-1 pretraining further improves convergence and final accuracy.}}
    \vspace{-5mm}
    \label{fig:ablation}
\end{figure*}

\textbf{Ablation on input representation and Stage-1 pretraining.} 
As shown in Figure~\ref{fig:ablation}, feat5+stage1 achieves the best performance, with 17.9° azimuth error / 7.70° elevation error / 1.13 distance error, substantially outperforming raw4+stage1 (30.5° / 13.8° / 1.65). This suggests that, compared with directly modeling the raw FOA waveform, the pre-extracted 5-channel acoustic features provide stronger localization cues. Moreover, comparing feat5 with feat5+stage1 shows that removing Stage-1 semantic alignment pretraining degrades performance, indicating that such initialization improves optimization and helps the model converge to a better solution. Further tests show that simply adding an FOA encoder, without Stage~1 and Stage~2 training, does not bring clear gains over the variant without the FOA encoder. This result highlights the necessity of dedicated pretraining to avoid interference during multimodal instruction tuning that jointly mixes planning and QA tasks.
\begin{table}[t]
    \centering
    \fontsize{8pt}{8.4pt}\selectfont
    \setlength{\tabcolsep}{3.5pt}
    \renewcommand{\arraystretch}{1.15}

    \caption{{\textbf{Ablation of the native audio encoder, FOA encoder, and Stage~3 training strategy on OmniEchoBench-QA.} The best overall performance is obtained by combining both encoders and freezing the FOA encoder during Stage~3.}}

    \label{tab:ablation}
    \vspace{-4mm}
    \scalebox{0.8}{
    \begin{tabular}{c c c c c c c c c}
        \toprule
        \makecell[c]{\textbf{Native audio encoder}} &
        \makecell[c]{\textbf{FOA encoder}} &
        \makecell[c]{\textbf{FOA encoder (stage 3)}} &
        \makecell[c]{\textbf{Camera}\\\textbf{Rotation}} &
        \makecell[c]{\textbf{Source}\\textbf{Direction}} &
        \makecell[c]{\textbf{3D}\\\textbf{Localization}} &
        \makecell[c]{\textbf{Source}\\\textbf{Motion}} &
        \makecell[c]{\textbf{Cognitive}\\\textbf{Map}} &
        \makecell[c]{\textbf{Overall}} \\
        \midrule
        \ding{51} & \ding{51} & training & 37.5 & 21.6 & 13.2& 19.5& 45.2 & 26.2\\
        \ding{51} & \ding{55} & frozen & \textbf{43.7} & 16.2 & \textbf{14.2}& 16.3& 21.8 & 19.8\\
        \ding{55} & \ding{51}  & frozen & 40.1 & 18.9 & 13.6 & 21.0 & 47.2 &  26.2 \\
        \ding{51} &  \ding{51} & frozen & 41.7 & \textbf{22.6} & \textbf{14.2} & \textbf{24.7} & \textbf{47.5} & \textbf{28.5}  \\
        \bottomrule
    \end{tabular}}
    \vspace{-4mm}
\end{table}

\section{Conclusion}
We introduce \textbf{OmniEchoBench}, a real-world  benchmark unified for spatial audio-visual perception and sound-guided navigation, together with a controllable spatial audio rendering pipeline for data generation. 
We further propose \textbf{OmniEcho}, an omni-modal model that integrates an FOA encoder with a pretrained semantic audio pathway to jointly capture auditory semantics and spatial information.
Experiments demonstrate that our model achieves state-of-the-art performance in spatial audio perception and navigation. Also, the performance of sound-guided navigation is close to that of traditional text-guided VLN methods. 
These results highlight the value of spatial audio for embodied scene reasoning and navigation, while identifying fine-grained spatial localization and distance estimation as key remaining challenges. 
We hope this work inspires future research on embodied agents that jointly reason about \emph{what} they perceive and \emph{where} it originates.

\subsection*{AI use statement}
In this work, we used generative AI tools for generating synthetic datasets. Specifically, generative AI assisted in constructing the simulated spatial audio--visual and sound-guided navigation training data, including the generation of synthetic scene descriptions, audio-visual media, sound events, and associated annotations. The corresponding data-generation and validation procedures are reported in the main text and Appendix~\ref{app:data-generation}.

We did not use generative AI tools for the other tasks with required disclosure, including developing the core scientific contributions, designing the research methodology or experiments, conducting experimental analysis, selecting or verifying references, or making scientific claims; the remaining required-disclosure tasks are not applicable to this work.

Additionally, we used generative AI tools to aid or polish writing, including language editing, grammar correction, readability improvements, and suggestions of alternative wording. We also used generative AI tools for research ideation or execution, specifically to assist with writing code. We have reviewed all AI-assisted work. All AI-assisted text was carefully edited and approved by the authors; AI-assisted code was inspected and tested for correctness; synthetic data were checked using the validation procedures described in the paper; and all technical content, claims, citations, and reported results were manually verified. We take responsibility for the final content of this work, including text, claims, code, data, annotations, and other artifacts produced with the aid of generative AI.

\bibliography{iclr2027_conference}
\bibliographystyle{iclr2027_conference}
\newpage
\appendix
\section{Appendix}

\subsection{Data Generation Details}
\label{app:data-generation}

\subsubsection{SPATIAL AUDIOVISUAL SIMULATION FOR DATA SYNTHESIS.}

\begin{figure}[h]
    \centering
    \includegraphics[width=\linewidth]{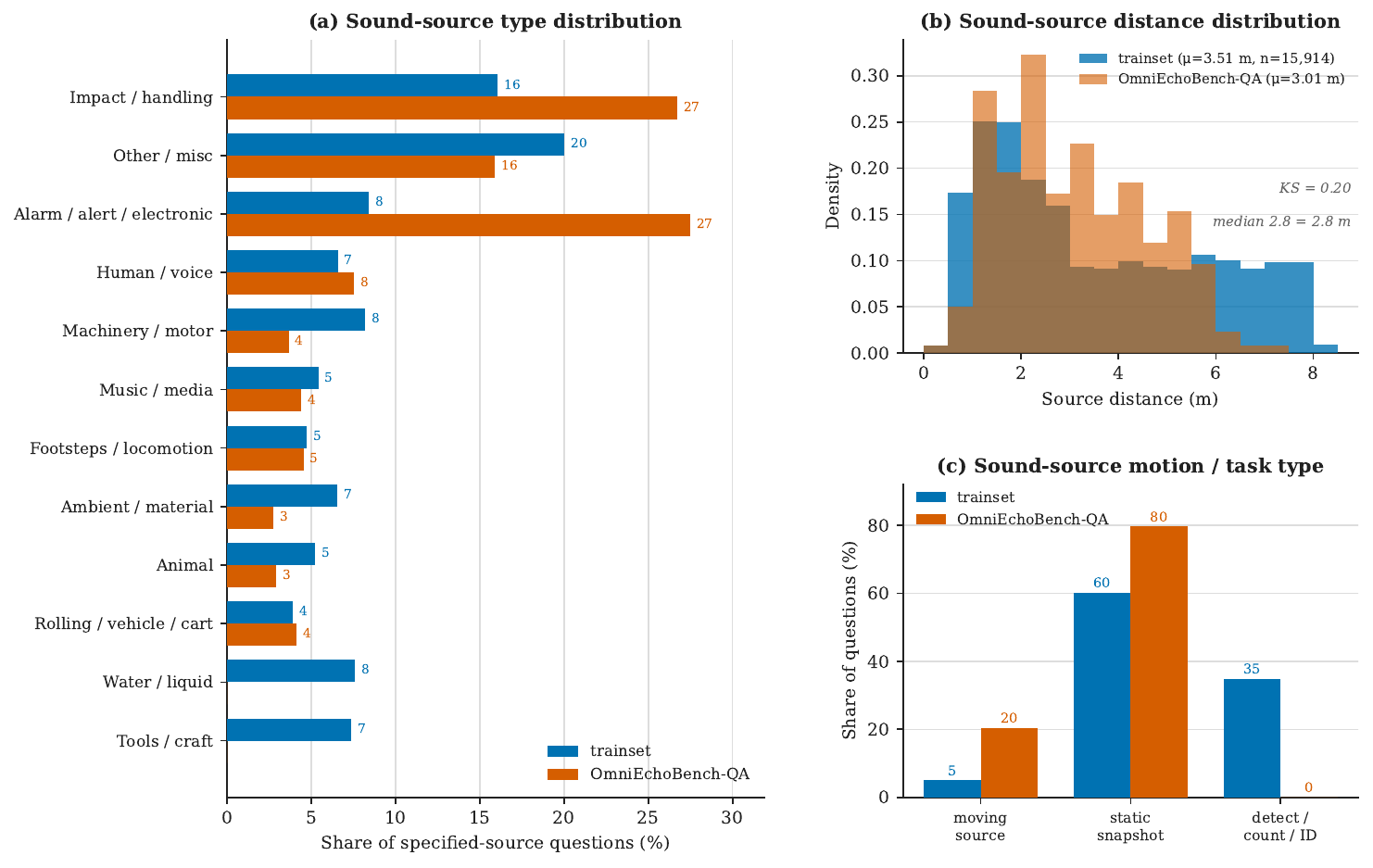}
    \caption{Distributional alignment between our synthetic training corpus
    (\textbf{trainset}, \num{145103} samples; blue) and the real-recorded
    \textbf{OmniEchoBench-QA} benchmark (orange), across three axes.
    \textbf{(a)}~\emph{Sound-source type.} Normalised over specified-source
    questions, the two corpora now share a comparable semantic mix
    ($\mathrm{JSD}=0.14$~bits); the residual gap is the benchmark's heavier
    emphasis on alarm/electronic tones, against the trainset's broader coverage of
    machinery, tools, and water sounds.
    \textbf{(b)}~\emph{Source distance.} The two distance
    distributions now largely overlap, with an identical median of
    \qty{2.8}{\meter} and a Kolmogorov--Smirnov statistic of $D=0.20$.
    \textbf{(c)}~\emph{Motion / task type.} We further augment the training set with non-spatial detect/count/identify tasks, aiming to improve the model's overall capability in spatial-audio perception.}

    \label{fig:dist-v3}
\end{figure}

We first construct spatial audiovisual QA data from generated videos. A large language model (LLM) acts as a \emph{director}: given previously generated scene descriptions to encourage diversity, it produces a structured JSON scene containing \(N\) foreground sources and one ambient source. For each source, the scene specifies a semantic label, target loudness (LUFS), onset and offset times, and a full trajectory \(\{(t_k,\theta_k,\phi_k,r_k)\}_k\). The same call also emits a full-scene video prompt describing all sounding entities and camera behavior, together with one isolated \emph{close-up} prompt per source that depicts only that sounding body in an otherwise silent environment. The director is driven by the system prompt shown in the box below (translated from the original and abridged; diversity, visibility-dynamics, and fine-grained grid/angle rules are elided with \texttt{[...]}). A set of auxiliary LLM agents subsequently refines the proposed scene---adjusting per-source onset/offset timing and target loudness, and resolving textual--geometric direction/distance contradictions---before any media are generated.

\begin{promptbox}[Scene director --- system prompt (translated, abridged)]
You are a Spatial Audio-Video Scene Director. Freely design one creative spatial-audio
scene and write a matching video-generation script (prompt).

The scene contains N_FOREGROUND foreground sources (e.g. a dripping tap, a cat, footsteps),
role="foreground", and exactly 1 ambient/background source (e.g. rain, wind, traffic),
role="ambient".

COORDINATES AND SPATIAL LAYOUT
- Right-handed frame; the listener faces +y.
  azimuth in [-180,180], 0 = straight ahead, +90 = left, -90 = right (+left / -right convention).
  elevation in [-90,90], + = up.  distance in [0.5m, 15m].
- The listener sits at the origin, position=[0,0,0].
- Each source is described by a trajectory (>=2 timestamped points); speed between adjacent
  points < 10 m/s.  Initial azimuths of different sources should be >= 30 deg apart.

CAMERA / LISTENER ORIENTATION
- The camera is bound to the listener; horizontal FOV ~= 90 deg (half-FOV 45 deg).
- listener.orientation.yaw is the camera azimuth and may vary over time via a trajectory.
- Visibility: camera_relative_azimuth(t) = object_azimuth(t) - listener_yaw(t), normalized to
  [-180,180]; the source is visible when |...| < 45 deg. Each source outputs a
  visibility_window {video_enter_t, video_exit_t}.

PHYSICAL-CONSISTENCY RULE FOR CAMERA ROTATION (critical)
- yaw increasing to positive => "camera rotates left"; decreasing to negative => "rotates right".
- When the camera turns toward X, the centered object must slide out toward not-X while a new
  object on the X side enters from the X edge to the center; violating this makes the
  video model produce reversed or frozen motion.  [... do/don't examples omitted ...]

OUTPUT FORMAT (strict) -- output ONE JSON object only, no markdown, no explanation:
{
  "scene": {
    "scene_id", "duration",
    "environment": {"type":"indoor|outdoor", "room":{"dimensions":[w,l,h],"rt60":0.3-0.8},
                    "description"},
    "listener": {"position":[0,0,0],
                 "orientation":{"yaw","pitch","roll","trajectory":[{"t","yaw"}, ...]}},
    "objects": [ {"id","type","label","role","loudness_lufs","snr_db","onset_t","offset_t",
                  "trajectory":[{"t","azimuth","elevation","distance"}, ...],
                  "visibility_window":{"video_enter_t","video_exit_t"}, "wav_path"} ]
  },
  "sub_prompts": [ "close-up video prompt per source (silent env, only that source sounding)", ...],
  "video_prompt": "full-scene prompt (3 parts: scene description; camera motion; METADATA appendix)"
}

LOUDNESS: near (<2m) foreground -14..-18 LUFS; mid (2-5m) -20..-26; far/ambient -30..-40;
  ~6 dB per distance doubling. snr_db is relative to the quietest (ambient=0 dB) source.
TIMING: at least one foreground onset_t=0; continuous/ambient sources span the whole clip.
SUB-PROMPTS: one close-up per source showing only that sounding body in an otherwise silent but
  reverberant environment; on-screen direction must match the source azimuth sign; static sources
  must be told to stay still.
DIVERSITY: vary environment (indoor/outdoor, city/nature/home/industrial, day/night, quiet/noisy)
  and source types across scenes; do not repeat similar scenes.  [... extensive diversity,
  dynamic-visibility, and grid/angle-annotation rules omitted ...]
\end{promptbox}

Before media generation, we validate the proposed scene for source counts, timing, spatial ranges, velocities, and visibility intervals. Invalid scenes are regenerated. We further compare directional and distance language in the prompts against the numerical trajectories, and a second LLM revises inconsistent descriptions. This checking--revision loop is repeated for up to three rounds. Additional planning adjusts source activity schedules and target loudness so that ambient sounds span the scene, transient events occur at plausible times, and foreground sounds remain audible over the background.

We then submit the prompts to Seedance~\citep{seedance2026seedance}. The full-scene prompt yields the rendered video \(V\), while each close-up prompt yields an isolated clip whose audio track is extracted as that source's monophonic stem \(s_i\). Stems are trimmed or loop-extended to the intended duration, loudness-normalized to their target LUFS, and gated or faded before rendering.

Because the director's trajectories are only approximately consistent with the generated pixels, we recover source geometry directly from the rendered video. Under a static-camera assumption verified by dense optical flow, objects are localized by open-vocabulary text-prompted segmentation and tracked across frames, and each sounding source is matched to a detected visual instance by label. For a detected center pixel \((c_x,c_y)\) in a frame of size \(W \times H\), we recover the viewing direction under a pinhole camera model. Let
\[
f=\tfrac{W/2}{\tan(\mathrm{FOV}/2)}
\]
denote the horizontal focal length and \(f_v\) the vertical focal length. We first compute normalized image-plane coordinates
\begin{equation}
  u = \frac{W/2-c_x}{f}, \qquad
  v = \frac{H/2-c_y}{f_v},
\end{equation}
and then recover azimuth and elevation as
\begin{equation}
  \theta = \operatorname{atan2}(u, 1),
  \qquad
  \phi = \operatorname{atan2}\!\big(v, \sqrt{1+u^2}\big).
  \label{eq:invproj}
\end{equation}
Monocular depth on key frames provides relative range estimates, which are anchored to the director's scene scale to obtain metric distance. The resulting trajectories are smoothed and resampled at a fixed rate. Sources never observed on screen retain their scripted trajectory but are forced outside the FOV; dynamically exiting sources are linearly extrapolated; and disagreements between detected instance counts and the script trigger instance splitting or masking. This yields a corrected scene in which every source has a per-timestep \((\theta,\phi,r)\) trajectory together with an explicit visibility state.

Starting from the corrected scene, we augment the sample with additional off-screen events and camera motion. On-screen sources are kept unchanged, while \(2\)--\(4\) off-screen sources are appended at azimuths guaranteed to remain outside the possibly rotating FOV. We also account for sources entering or leaving the image and simulate their off-screen continuation. Each appended source follows a world-frame trajectory chosen from a static location, a straight walk, or a polyline with a single turn of at least \(90^\circ\), and is voiced by a borrowed stem from a cross-sample library after loudness normalization. A camera yaw trajectory \(\psi(t)\) is written into the scene so that the rendered audio rotates consistently with the cropped or panned image.

Given the final augmented scene, all stems are rendered into FOA using Eq.~\eqref{eq:foa}, with optional indoor reflections, and the resulting multichannel audio is muxed into the cropped or panned video without re-encoding the visual stream. We then generate QA pairs directly from the final trajectories rather than from an LLM response. At each queried time, we compute a source's world-frame direction, its direction relative to the current camera, its distance, and whether it lies inside the camera frustum; over intervals, we compute changes from the same trajectories. From these quantities we instantiate four categories of questions---direction (coarse 8-way and fine-grained degrees), motion (world heading and change), 3D localization, and camera rotation---covering eight subtypes across camera, camera-body, and world coordinate frames. Each video contributes \(15\)--\(20\) questions, with source and time selection balanced to cover both binary outcomes and the full set of direction labels. This process yields 26,585 training questions from 1,484 synthesized videos and the data distribution is shown in Figure~\ref{fig:dist-v3}.

\paragraph{{Audio-only Spatial Data}}
{We also synthesize spatial-audio questions without visual input. Each scene contains one to three independently generated sounds. We assign every source a time interval and either a static, constant-speed, accelerating, or decelerating 3D trajectory, then render the resulting mixture as FOA. Because the trajectories are known, we can directly ask about direction, elevation, distance, source count, motion, speed change, approach or recession, closest approach, and relations between sources. An LLM rewrites the questions and expresses the trajectory-based reasoning in natural language, while the answers remain fixed by the rendered scene. We obtain 18,518 questions from 2,000 moving-source scenes.}

The same single-source recordings are reused to construct 100,000 simpler static scenes for spatial-audio pretraining. Each clip contains one to four sound categories placed at independently sampled azimuths, elevations, and distances. These clips support the semantic-alignment and query-conditioned localization stages of the FOA encoder described in Section~4.

\subsubsection{Sound-Guided Navigation Data Synthesis}
\label{app:nav-synth}

We convert four existing VLN corpora---R2R, RxR, VLN-PE, and ScaleVLN---into sound-guided navigation supervision through a six-stage pipeline: (i) destination sound-event proposal, (ii) audio synthesis, (iii) FOA rendering and sample construction, (iv) acoustic augmentation, (v) omni-modal SFT formatting, and (vi) chain-of-thought (CoT) annotation.

\textbf{Destination sound-event proposal.} For each episode, a vision--language model (Qwen3.7-flash) observes the last five first-person frames captured at the destination (front camera, $1.25$\,m height) and proposes a plausible sound event, under a hard constraint that neither the sound content nor the derived task may reveal the destination location, room name, or nameable furniture, so that the agent must localize the target acoustically. Proposals are balanced roughly $50/50$ between two categories: (1) a short \emph{spoken utterance} (a person calling the robot, with diverse intents and an accompanying voice-style description of gender, age, timbre, emotion, and pace), and (2) a \emph{non-speech ambient event} (e.g., knocking, running water, dragging furniture, an animal, a phone, or an appliance). The model returns a structured record containing the sound content, a natural-language task description, the voice/acoustic style, and its scene reasoning. The system prompt is reproduced verbatim below.

\begin{promptbox}[Destination sound-event proposal --- system prompt]
You are a scenario designer for an indoor "sound-source navigation" task. A robot starts
somewhere in a house and must reach a specific target object/location ONLY by following a sound
emitted from that target (plus a short text instruction). You are shown the LAST few first-person
camera frames of a navigation episode; these frames show the destination area (the target). Your
job: invent a plausible sound event that would emit from this destination, so the robot can
localize it acoustically.

Design two categories of scripts; pick ONE per episode (respect the requested category when
given, and keep the two categories roughly balanced across the dataset):
  Category 1 - "spoken utterance that calls the robot over": the sound itself is a short spoken
    English utterance from a person at the destination. This is NOT limited to emergencies or
    calls for help - ANY natural reason a person would summon the robot is valid, and you should
    keep them DIVERSE across episodes, e.g.: asking for help ("Robot, I fell down, please come and
    find me."), requesting an item be fetched or carried, inviting it to come look at something,
    assigning a task, a casual/curious call, a playful summon, an impatient shout, etc. Vary the
    speaker, wording, tone and intent widely. The text_instruction can briefly restate the intent
    (e.g. "Go to the person who is calling you.").
    For Category 1 you MUST also fill "voice_style": one English sentence describing the SPEAKER'S
    VOICE for downstream TTS - gender, approximate age, timbre, emotion/tone, speaking pace, and
    optionally accent. Make these rich and varied across episodes.
  Category 2 - "ambient sound source + text command": the sound is a common non-speech sound
    believable in this room (knocking, running water / flushing, dragging furniture, a pet - dog
    barking / cat meowing, human chatter, music, TV, kettle whistle, phone ringing, etc.). The
    text_instruction is a short English command telling the robot what to home in on, e.g. "Find
    where the knocking sound is coming from." For Category 2, set "voice_style" to a short English
    description of the sound's acoustic character or "" if not applicable.

Use the visible scene ONLY to make the sound realistic (bathroom -> water/flush; kitchen ->
clattering dishes / kettle; living room -> TV / music; bedroom -> alarm clock / phone).

HARD CONSTRAINTS:
- NEVER reveal the destination's location, room name, or nameable furniture in either
  sound_content or text_instruction. The robot must find it by sound, not by being told where it
  is. (Do NOT say things like "in the bathroom" / "near the toilet".)
- sound_content, text_instruction and voice_style must be in ENGLISH.
- Keep each field to one short natural sentence.

Return ONLY a JSON object, no markdown fences, with exactly these keys:
{
  "category": 1 or 2,
  "sound_content": "<the audio content: cat.1 = the spoken line; cat.2 = a short sound description>",
  "text_instruction": "<short English instruction given to the robot>",
  "voice_style": "<cat.1 = the speaker's voice; cat.2 = the sound's acoustic character or empty>",
  "scene_reasoning": "<one short sentence: why this sound fits what you see>"
}
\end{promptbox}

\begin{table}[h]
    \centering
    \fontsize{7.3pt}{9.2pt}\selectfont
    \setlength{\tabcolsep}{5pt}
    \renewcommand{\arraystretch}{1.15}
    \caption{Audio synthesis models used for sound-guided navigation data.}
    \label{tab:app-nav-tts}
    \begin{tabular}{@{}l l l@{}}
        \toprule
        \textbf{Sound category} & \textbf{Synthesizer} & \textbf{Key settings} \\
        \midrule
        Spoken utterance & Custom-voice TTS (Qwen3-TTS)~\citep{qwen3tts2026} & 7 speakers (4 F / 3 M), style-conditioned, $\le2048$ tokens \\
        Non-speech event & Seedance~\citep{seedance2026seedance} & clip $10$\,s \\
        \bottomrule
    \end{tabular}
\end{table}

\begin{table}[h]
    \centering
    \fontsize{8pt}{9.2pt}\selectfont
    \setlength{\tabcolsep}{5pt}
    \renewcommand{\arraystretch}{1.15}
    \caption{Rendering and augmentation configuration for sound-guided navigation data.}
    \label{tab:app-nav-render}
    \begin{tabular}{@{}l l@{}}
        \toprule
        \textbf{Component} & \textbf{Setting} \\
        \midrule
        Source placement & episode endpoint (center of final destination view) \\
        Distance attenuation & inverse distance $\propto 1/r$, floor $r_{\min}=0.3$\,m, no peak normalization \\
        FOA format & first-order ambisonics, ACN/SN3D, 4 channels, order $[W,Y,Z,X]$ \\
        Listener / camera height & $1.25$\,m \\
        History frames & up to $8$ (uniformly sampled) \\
        Waypoints & $K=8$, $[x_{\text{fwd}},\,y_{\text{left}},\,\psi]$ relative to current pose \\
        Room reverberation & R2R / RxR: image-source early reflections + cross-room attenuation (MP3D) \\
                            & ScaleVLN / VLN-PE: direct-path propagation only \\
        Directional noise (optional) & $1$--$3$ plane-wave interferers (pink/white); el.\ $\pm45^\circ$; $\ge30^\circ$ from source \\
        Noise SNR & sampled from $\{20,15,10,5,0\}$\,dB w.r.t.\ the $W$ channel \\
        \bottomrule
    \end{tabular}
\end{table}

\textbf{Audio synthesis.} Spoken utterances are synthesized with a custom-voice text-to-speech model conditioned on the voice-style description, and non-speech events with Seedance, as summarized in Table~\ref{tab:app-nav-tts}.

\textbf{FOA rendering and sample construction.} The proposed source is placed at the episode endpoint (the center of the final destination view). At every decision frame, we transform the endpoint into the listener's current camera frame to obtain its forward/left/up direction components and distance $r$; the mono stem is inverse-distance attenuated ($\propto\!1/r$, floored at $r_{\min}=0.3$\,m, with no per-clip peak normalization so that loudness preserves a distance cue) and encoded into 4-channel FOA via Eq.~\eqref{eq:foa}. Trajectory targets are the next $K=8$ waypoints $[x_{\text{fwd}},\,y_{\text{left}},\,\psi]$ (forward metres, left metres, yaw radians) expressed relative to the current pose. Key constants are listed in Table~\ref{tab:app-nav-render}.

\textbf{Acoustic augmentation.} For corpora with room geometry (R2R and RxR on Matterport3D), the shared renderer adds first-order image-source early reflections and attenuates direct sound across rooms~\citep{allen1979image}, and RxR further adds low-level directional background noise; geometry-free corpora (ScaleVLN on HM3D, VLN-PE) use direct-path propagation with distance attenuation only. On top of this, we provide an optional directional-noise augmentation that injects $1$--$3$ plane-wave interferers (pink or white noise) at random azimuths (elevation within $\pm45^\circ$, at least $30^\circ$ from the true source direction), mixed at a signal-to-noise ratio sampled per clip from $\{20,15,10,5,0\}$\,dB referenced to the omnidirectional ($W$) channel (Table~\ref{tab:app-nav-render}).

\textbf{Omni-modal SFT format and CoT.} Each SFT example interleaves up to eight uniformly sampled history frames, the current frame, and the co-located 4-channel FOA clip, followed by the task text; the target is the $K=8$ waypoint list. A vision--language model then writes a first-person CoT rationale conditioned on the ground-truth source bearing (bucketed into eight directions with $22.5^\circ/67.5^\circ/112.5^\circ/157.5^\circ$ edges), distance (near $<1.5$\,m, far $>4$\,m), and two trajectory-derived meta-actions, without altering the trajectory targets. We additionally construct in-place turning examples (sources to the side or behind the agent). Active-stop supervision converts stationary end-of-episode waypoints into a fixed eight-slot trajectory carrying a majority of \texttt{<stop>} tokens ($\ge5$ of $8$). The eight-way direction bucket, the near/far distance bucket, and the two trajectory-derived meta-actions are computed deterministically from geometry and injected into the CoT prompt as ground truth; the prompt template is shown below.

\begin{promptbox}[Navigation chain-of-thought --- system and task prompt]
[SYSTEM]
You are writing first-person chain-of-thought (CoT) reasoning traces AS a sound-source navigation
robot. You (the robot) hear a spatial-audio cue and must move toward the sound source while
avoiding obstacles. You are given YOUR OWN recent egocentric RGB frames (oldest to newest, the
last one is your current view), the task the sound is asking you to do, the GEOMETRICALLY COMPUTED
direction and distance of the sound source relative to you, the ground-truth meta-action sequence,
and the ground-truth future trajectory. Write a single coherent, physically plausible reasoning
trace, STRICTLY IN THE FIRST PERSON (use 'I', 'me', 'my', 'ahead of me'; NEVER say 'the robot').
The computed direction/distance and the meta-action sequence are the SOURCE OF TRUTH: never
contradict them. Describe the visual scene ONLY from what you actually see in the frames.

[TASK]
Write a first-person chain-of-thought reasoning trace for my next navigation step.

=== Task the sound is asking me to do ===
{instruction}

=== Computed sound-source cue (ground truth -- do NOT contradict) ===
- Direction of the sound source relative to me: {direction}  (azimuth {az:+.0f} deg, where + is to
  my left, 0 is straight ahead)
- Distance: {dist:.1f} m ({dist_bucket})
- Elevation: {el:+.0f} deg (roughly at my ear level unless large)

=== Ground-truth motion (derived from the future trajectory -- do NOT contradict) ===
- My two main meta-actions (first half of my path, then second half): {meta_seq}
- Net motion over this horizon: {net_fwd:+.2f} m forward, {net_left:+.2f} m to my left; my heading
  changes by {net_yaw:+.0f} deg.
- Overall travel direction of my path: {path_dir}
- {deviation_note}

=== What to write ===
Produce ONE first-person reasoning trace (a few sentences, natural English, no bullet headers, no
markdown) that flows through these stages IN ORDER:
1. SCENE: briefly describe what I see in front of me / around me in my current view.
2. INSTRUCTION: restate, in one clause, what the sound is asking me to do.
3. SOUND DIRECTION: state where I hear the sound source ({direction}, {dist_bucket}) and tie it to
   what I see.
4. PLAN: give a short high-level plan that reconciles what I see (obstacles/openings) with where I
   hear the sound. If my path direction differs from the straight line to the source, explain it as
   going around an obstacle or leaving my current room; if the source is far, consider it may be in
   another room.
5. DECISION: conclude in ONE short first-person sentence that commits to my two sequential moves --
   what I do over the first half of my path, then the second half -- consistent with: {meta_seq}.
   Do NOT write a "Meta actions:" label, do NOT enumerate sub-moves, and do NOT output coordinates.

Write everything in the FIRST PERSON. Do NOT refer to "the robot" in the third person. Do NOT
output the numeric trajectory or any coordinates. Stop at the meta-action level.

Output STRICT JSON and nothing else, exactly:
{"cot": "<the first-person reasoning trace as a single string>"}
\end{promptbox}

\subsection{Benchmark Details}

\subsubsection{OmniEchoBench-QA: Real-World Capture and Annotation}
\label{app:bench-qa}
\begin{figure*}[t]
    \centering
    \includegraphics[width=\textwidth]{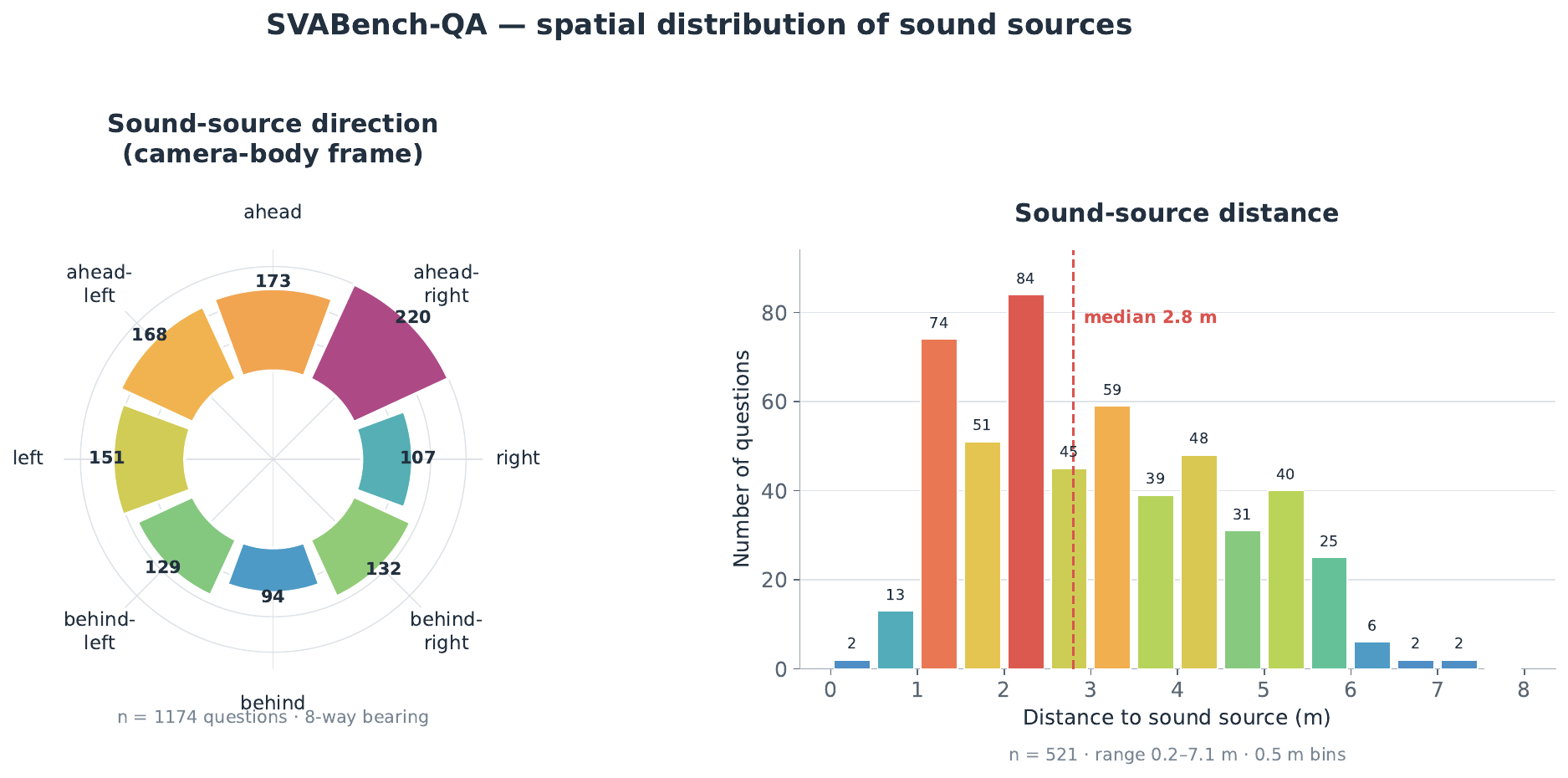}
    \caption{{Direction and Distance distribution of OmniEchoBench-QA.}}
    \label{fig:qa-distribution}
\end{figure*}

\begin{figure*}[t]
    \centering
    \includegraphics[width=\textwidth]{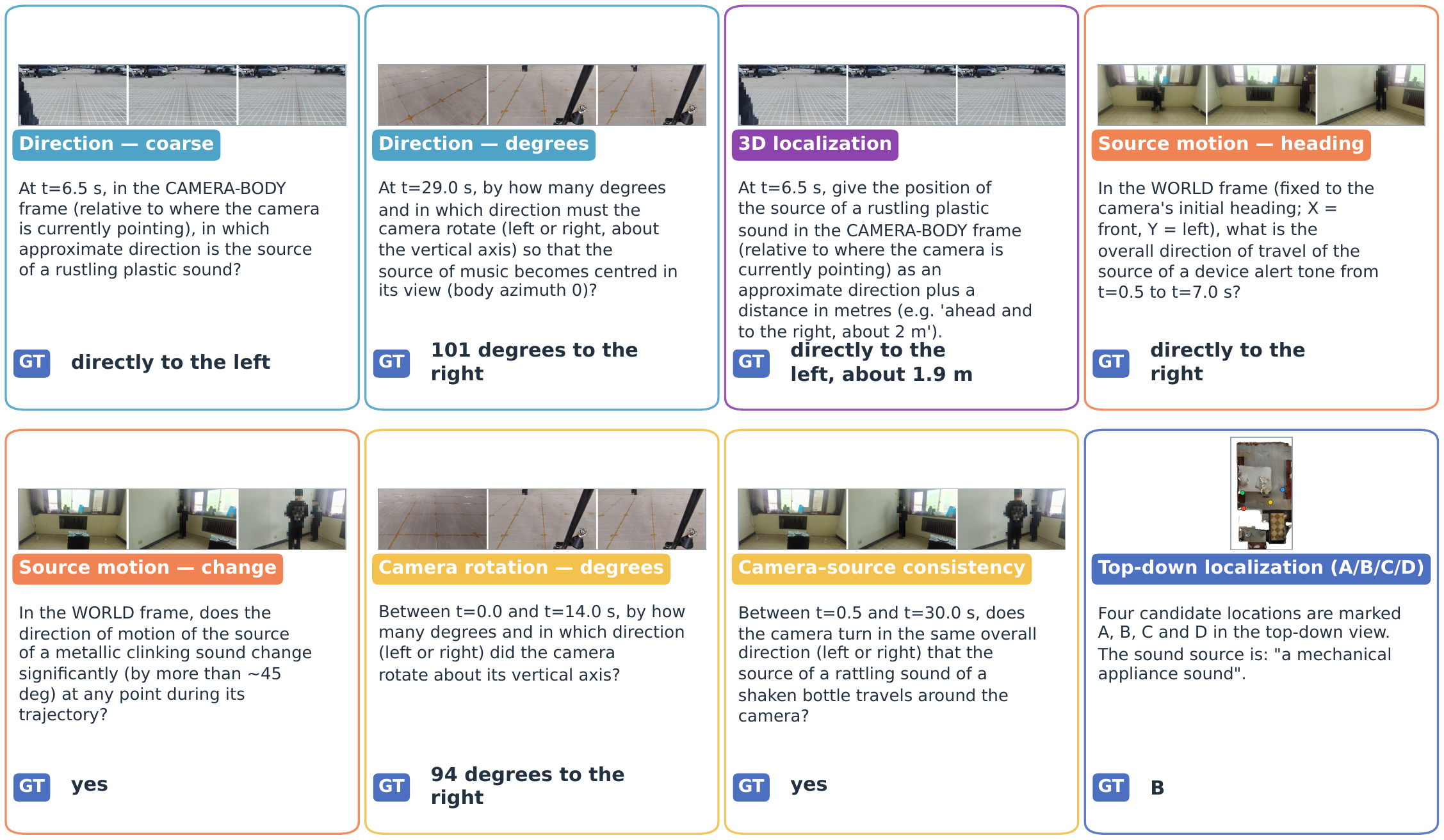}
    \caption{Representative examples from OmniEchoBench-QA, covering eight spatial reasoning question types.}
    \label{fig:qa-all-types-vis}
\end{figure*}

\textbf{Capture rig and formats.} Each script is performed and recorded live, producing three synchronized streams in a single take: a first-person main-view video ($1920$\,px wide, $30$\,fps), a $360^\circ$ panoramic video, and a four-channel FOA audio track recorded at $48$\,kHz. The FOA audio is stored with the omnidirectional channel first, and its horizontal axes are aligned so that the front direction corresponds to the $+X$ axis and the listener's left to the $+Y$ axis. The $197$ scripts are distributed across small indoor ($50$), medium indoor ($50$), large indoor ($77$), and open outdoor ($20$) settings.

\textbf{Annotation.} Using the panoramic recording as ground-truth reference, annotators label the real-time position of every active sound source at $2$\,Hz (every $0.5$\,s), together with per-timestep flags for whether the source is visible in the main view, visible in the panorama, and audible. The panorama supplies the "present but off-screen" evidence that the main view cannot, and is used only during annotation. Two capture regimes are used: a \emph{camera-fixed} regime (the original FOA is kept) and a \emph{camera-moving} regime, in which the camera yaw is recovered by optical flow and calibrated against annotated rotation anchors, and the FOA is rotated into the current camera frame so that audio and image remain aligned. At test time, the model receives the main-view video (or a single frame) muxed with the processed four-channel FOA.

\textbf{Question construction.} All ground-truth answers are computed deterministically from the annotated trajectories rather than from a language model: at a queried time we compute the source's world and camera-body bearing, its distance, and its visibility; over an interval we compute changes from the same trajectories. Answers are short phrases drawn from closed buckets (eight-way direction words, \emph{$N$ degrees to the left/right}, yes/no, or a direction plus a metric distance), and per-(category, subtype) answer-bucket balancing is applied. The resulting temporal-reasoning categories and counts are listed in Table~\ref{tab:app-qa-cats}. The main paper reports the four geometric categories (Direction, 3D~Localization, Motion, Camera~Rotation; $2{,}372$ QA); a Visibility category is additionally annotated and available. The distribution of direction and distance is shown in Figure~\ref{fig:qa-distribution}. The Cognitive~Map subset ($600$ QA) is built on the navigation scenes below. Example QAs from OmniEchoBench-QA are shown in Figure~\ref{fig:qa-all-types-vis}.

\textbf{Coordinate conventions.} The world frame is fixed to the camera's initial heading, with $+X$ ahead and $+Y$ to the left; the camera-body frame at time $t$ rotates with the recovered camera yaw $\theta(t)$, and body azimuth equals world azimuth minus $\theta(t)$. Azimuth is measured counter-clockwise ($0^\circ$ ahead, $90^\circ$ left, $180^\circ$ behind, $270^\circ$ right).

\begin{table}[t]
    \centering
    \fontsize{8pt}{9.2pt}\selectfont
    \setlength{\tabcolsep}{5pt}
    \renewcommand{\arraystretch}{1.15}
    \caption{OmniEchoBench-QA temporal-reasoning categories, subtypes, coordinate frames, and counts (from \texttt{spatial\_qa\_test}). The four geometric categories ($2{,}372$ QA) are reported in the main paper; Visibility is additionally annotated.}
    \label{tab:app-qa-cats}
    \begin{tabular}{@{}l l l c@{}}
        \toprule
        \textbf{Category} & \textbf{Subtype(s)} & \textbf{Frame} & \textbf{Count} \\
        \midrule
        Source Direction & approx.\ 8-way / degrees & camera-body & $1{,}069$ \\
        3D Localization & bearing $+$ distance & camera-body & $521$ \\
        Source Motion & world heading / change & world & $473$ \\
        Camera Rotation & degrees / consistency & camera / mixed & $309$ \\
        \bottomrule
    \end{tabular}
\end{table}

\subsubsection{OmniEchoBench-Nav: Real-World Scene and Acoustic-Field Collection}
\label{app:bench-nav}

\textbf{Scenes.} The benchmark contains $30$ real-scanned indoor environments, split into $10$ single-room and $20$ multi-room (cross-room) layouts spanning two to four connected rooms. Each scene is provided as a textured mesh (\texttt{.glb}) together with a top-down render and a pixel-to-mesh affine calibration; walkability, collision checks, and shortest-path planning are provided by an occupancy-grid navmesh derived from the mesh. 

\textbf{Acoustic field.} Each scene contains $30$ navigation samples ($900$ in total). In every sample a single mono source is placed at a known $3$D position, and its signal is captured at a dense grid of $12$--$17$ receiver ("microphone") positions on an approximately $1$\,m grid at $\sim\!1$\,m height, each stored as a world-aligned four-channel FOA recording (ACN/SN3D, order $[W,Y,Z,X]$, $48$\,kHz), forming a spatially dense acoustic field of the environment. Every receiver's coordinates are stored explicitly. Sources span non-speech environmental events ($505$ samples) and spoken commands ($395$ samples), across a set of playback devices; each sample is annotated with the source device, source room, semantic description, spoken-command text (when applicable), speaker description, and a natural-language navigation-target description. Table~\ref{tab:app-nav-bench} summarizes the composition.

\textbf{Test-time perception.} At each step, the agent is served the FOA recording from the receiver nearest to its current position. Although the recordings are stored in a world-aligned coordinate frame, their horizontal directional components are rotated into the agent's current egocentric frame according to the camera yaw before being passed to the model. The visual observation is rendered from the same pose and orientation ($+x$ forward, $90^\circ$ horizontal FOV, $15^\circ$ downward pitch), ensuring that the FOA forward direction remains aligned with the camera view as the agent turns. The Cognitive~Map QA subset ($600$ four-option A/B/C/D questions) is built on these scenes: it pairs the egocentrically aligned nearest-receiver FOA with a top-down map on which the true source and three distractor floor points (each $\ge40^\circ$ apart in bearing) are marked, and asks which marked location is the source. In the ego-up variant the map is rotated so image-up is the robot's forward direction and the robot is centered.

\begin{table}[t]
    \centering
    \fontsize{8pt}{9.2pt}\selectfont
    \setlength{\tabcolsep}{6pt}
    \renewcommand{\arraystretch}{1.15}
    \caption{OmniEchoBench-Nav composition. $30$ scenes $\times$ $30$ samples $= 900$ navigation samples; each sample provides a dense FOA receiver grid.}
    \label{tab:app-nav-bench}
    \begin{tabular}{@{}l l@{}}
        \toprule
        \textbf{Property} & \textbf{Value} \\
        \midrule
        Scenes & $30$ ($10$ single-room, $20$ multi-room / $2$--$4$ rooms) \\
        Navigation samples & $900$ ($30$ per scene) \\
        Receivers per sample & $12$--$17$ (dense $\sim\!1$\,m grid, $\sim\!1$\,m height) \\
        Source types & non-speech $505$, spoken command $395$ \\
        FOA format & 4-channel, ACN/SN3D $[W,Y,Z,X]$, $48$\,kHz, world-aligned \\
        Scene assets & textured \texttt{.glb} mesh $+$ top-down render $+$ occupancy navmesh \\
        Cognitive-Map QA & $600$ four-option (A/B/C/D) questions over all $30$ scenes \\
        \bottomrule
    \end{tabular}
\end{table}

\subsubsection{Data Privacy and De-identification}
\label{app:data-privacy}

Because OmniEchoBench contains audiovisual recordings captured in real-world environments, we process the data to minimize the disclosure of personally identifiable information before any sample is included in the benchmark or prepared for release. In all visual streams, every visible face and human figure, including incidental bystanders, is mosaic-blurred. Only these de-identified visual streams and privacy-preserving audio recordings are included in the benchmark and prepared for release. The released data are intended for research on spatial audio--visual perception and navigation, not for identity recognition, person re-identification, speaker identification, or other biometric analysis.

\subsection{Training Details}
\label{app:training-details}

OmniEcho is trained in three stages: two stages pretrain the standalone FOA encoder (semantic alignment, then query-conditioned localization), and a third stage integrates the frozen encoder into the Qwen3-Omni backbone via a trainable projector. The FOA encoder is a compact ($\sim\!11.9$\,M-parameter) convolution--Transformer network; its architecture is summarized in Table~\ref{tab:app-enc-arch}.

\begin{table}[t]
    \centering
    \fontsize{8pt}{9.2pt}\selectfont
    \setlength{\tabcolsep}{6pt}
    \renewcommand{\arraystretch}{1.15}
    \caption{FOA encoder architecture ($\sim\!11.9$\,M parameters). The encoder is shared across all three training stages and frozen at Stage~3.}
    \label{tab:app-enc-arch}
    \begin{tabular}{@{}l l@{}}
        \toprule
        \textbf{Module} & \textbf{Configuration} \\
        \midrule
        Input feature map & $5\times128\times T$: log-mel of $W$ $+$ DirAC intensity $(i_x,i_y,i_z)$ $+$ diffuseness \\
        STFT front-end & $16$\,kHz, FFT $400$, hop $160$, $128$ mel bins \\
        Conv.\ frontend & $3\times$ Conv2d (stride $2$, GELU), $8\times$ time downsampling ($\sim\!12.5$ tokens/s) \\
        Transformer & $5$ layers, $d_{\text{model}}=384$, $6$ heads, FFN $1536$, pre-norm, sinusoidal PE \\
        Semantic head (Stage 1) & learnable-query attention pooling $\rightarrow$ $512$-d clip embedding \\
        Text encoder (frozen) & CLIP ViT-B/16~\citep{radford2021clip}, $512$-d joint space, $C=644$ sound labels \\
        Localization head (Stage 2) & query-conditioned cross-attention $\rightarrow$ $(\text{az},\text{el},\text{dist})$ \\
        \bottomrule
    \end{tabular}
\end{table}
\paragraph{Stage 1: Semantic alignment of the FOA encoder.}
We first pretrain a lightweight FOA audio encoder $f_\theta$ that maps an FOA audio clip to a temporal sequence of $d$-dimensional tokens ($d=384$). Each clip is converted into a 5-channel input map $\mathbf{X}\in\mathbb{R}^{5\times128\times T}$ comprising the log-mel spectrogram of the omnidirectional ($W$) channel together with the three active-intensity components $(i_x,i_y,i_z)$ and the diffuseness $\delta$, all in the ACN/SN3D convention. A convolutional frontend downsamples time by $8\times$ (yielding ${\sim}12.5$ tokens/s), and a 5-layer Transformer produces token embeddings. An attention-pooling semantic head aggregates these tokens into a clip embedding $\mathbf{a}\in\mathbb{R}^{512}$.

To inherit an open-vocabulary sound-semantic space, we align the clip embedding with frozen CLIP text embeddings using the balanced sigmoid objective defined below. This balancing is important because positive audio--label pairs constitute only ${\sim}0.3\%$ of all clip--label pairs in our training setup. Stage~1 is trained on a $100$k-clip corpus of synthetic static FOA scenes ($1$--$4$ sources, $2$--$8$\,s), rendered with an order-1 spherical-harmonic encoder from mono stems at known azimuth, elevation, and distance.

\paragraph{Balanced semantic alignment objective.}
Let $\mathbf{a}_i\in\mathbb{R}^{512}$ denote the audio embedding of clip $i$, and let $\mathbf{t}_j\in\mathbb{R}^{512}$ be the frozen CLIP text embedding of label $j$, where $j\in\{1,\dots,C\}$ and $C=644$. Before similarity computation, both embeddings are $\ell_2$-normalized:
\begin{equation}
\tilde{\mathbf a}_i=\frac{\mathbf a_i}{\|\mathbf a_i\|_2},
\qquad
\tilde{\mathbf t}_j=\frac{\mathbf t_j}{\|\mathbf t_j\|_2}.
\end{equation}
We then form sigmoid logits as
\begin{equation}
s_{ij}=\alpha\,\tilde{\mathbf a}_i^{\!\top}\tilde{\mathbf t}_j+b,
\qquad
\alpha=\exp(\gamma),
\end{equation}
where $\alpha$ is a learnable logit scale and $b$ is a learnable bias. For clip $i$, let $\mathcal{P}_i\subseteq\{1,\dots,C\}$ be the set of present labels and let $\mathcal{N}_i=\{1,\dots,C\}\setminus\mathcal{P}_i$ be the set of absent labels. We use a per-clip class-balanced sigmoid loss,
\begin{equation}
\mathcal{L}_{\mathrm{sem}}^{(i)}
=
\frac{1}{2}\left(
\frac{1}{|\mathcal{P}_i|}\sum_{j\in\mathcal{P}_i}\mathrm{BCE}(s_{ij},1)
+
\frac{1}{|\mathcal{N}_i|}\sum_{j\in\mathcal{N}_i}\mathrm{BCE}(s_{ij},0)
\right),
\end{equation}
and average it over the mini-batch:
\begin{equation}
\mathcal{L}_{\mathrm{sem}}
=
\frac{1}{B}\sum_{i=1}^{B}\mathcal{L}_{\mathrm{sem}}^{(i)}.
\end{equation}
Here
\begin{equation}
\mathrm{BCE}(s,y)
=
-\big[y\log\sigma(s)+(1-y)\log(1-\sigma(s))\big].
\end{equation}
The same $\mathcal{L}_{\mathrm{sem}}$ is used in both Stage~1 and Stage~2.

\paragraph{Stage 2: Query-conditioned spatial localization.}
Initialized from Stage~1, the encoder is jointly fine-tuned for sound-source localization. Given a text query embedding $\mathbf{q}$ describing one source present in the clip, a localization head attends over the encoder tokens via query-conditioned cross-attention and regresses the source direction and range. Rather than predicting azimuth $a$ and elevation $e$ directly, the head emits their $(\sin,\cos)$ encodings together with a log-distance, i.e., a raw 5-vector $(\hat{s}_a,\hat{c}_a,\hat{s}_e,\hat{c}_e,\hat{\ell})$. The two angular pairs are $\ell_2$-normalized before the loss is applied, which avoids the $\pm 180^\circ$ wrap-around discontinuity.

Stage~2 minimizes a localization regression loss while retaining the same semantic objective $\mathcal{L}_{\mathrm{sem}}$ as a regularizer to prevent semantic drift:
\begin{equation}
\mathcal{L}_{\mathrm{stage2}}
=
\mathcal{L}_{\mathrm{ang}}
+
\lambda_d\,\mathcal{L}_{\mathrm{dist}}
+
\lambda_s\,\mathcal{L}_{\mathrm{sem}},
\qquad
\lambda_d=\lambda_s=0.5.
\end{equation}

The angular term is the mean-squared error between the normalized predicted $(\sin,\cos)$ pairs and the ground-truth pose $(a,e)$:
\begin{equation}
\mathcal{L}_{\mathrm{ang}}
=
\Big\| \widehat{\mathbf{u}}_a - (\sin a,\cos a) \Big\|_2^2
+
\Big\| \widehat{\mathbf{u}}_e - (\sin e,\cos e) \Big\|_2^2,
\end{equation}
with
\begin{equation}
\widehat{\mathbf{u}}_a
=
\frac{(\hat{s}_a,\hat{c}_a)}{\|(\hat{s}_a,\hat{c}_a)\|_2},
\qquad
\widehat{\mathbf{u}}_e
=
\frac{(\hat{s}_e,\hat{c}_e)}{\|(\hat{s}_e,\hat{c}_e)\|_2}.
\end{equation}

The range is supervised in the log domain with an $\ell_1$ term against the ground-truth distance $r$:
\begin{equation}
\mathcal{L}_{\mathrm{dist}}
=
\big|\,\hat{\ell}-\log r\,\big|.
\end{equation}

Both encoder stages are trained on the same $100$k-clip synthetic FOA corpus; the optimization settings are listed in Table~\ref{tab:app-enc-hparams}.

\begin{figure*}[t]
    \centering
    \includegraphics[width=\textwidth]{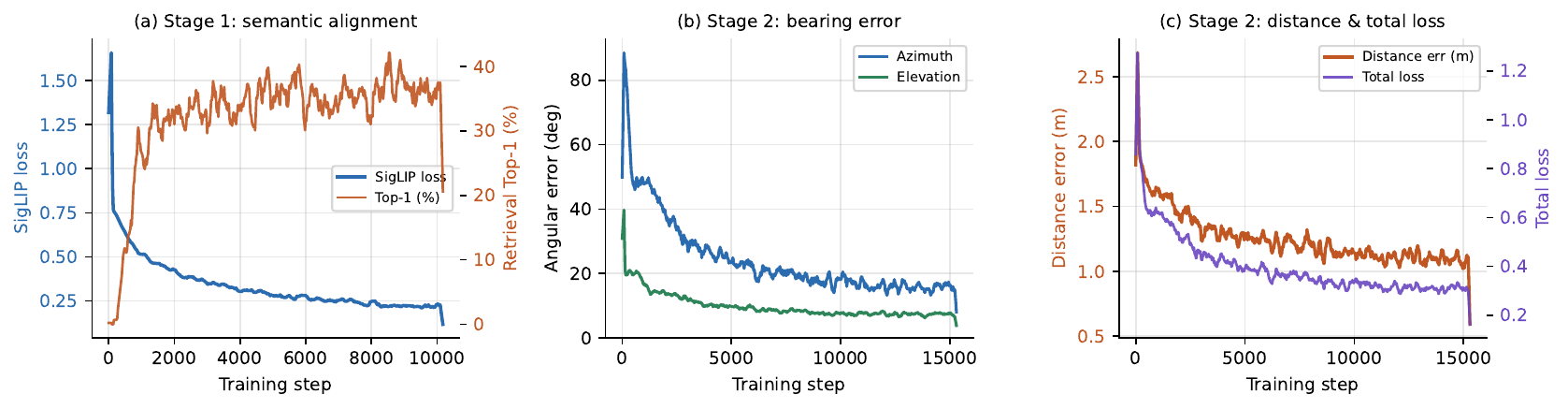}
    \caption{FOA encoder pretraining curves (smoothed). (a) Stage~1 semantic alignment: SigLIP loss and open-vocabulary retrieval Top-1 accuracy. (b) Stage~2 query-conditioned localization: mean azimuth and elevation error. (c) Stage~2 distance error and total loss.}
    \label{fig:training-curves}
\end{figure*}

Figure~\ref{fig:training-curves} shows the encoder pretraining dynamics. In Stage~1, the semantic loss drops to $\sim\!0.23$ while open-vocabulary retrieval Top-1 rises to $\sim\!37\%$ over 644 labels; in Stage~2, the mean azimuth error falls to $\sim\!16^\circ$, elevation error to $\sim\!7^\circ$, and distance error to $\sim\!1.1$\,m, confirming that the compact encoder becomes both semantically grounded and spatially discriminative before it is frozen.

\begin{table}[t]
    \centering
    \fontsize{8pt}{9.2pt}\selectfont
    \setlength{\tabcolsep}{8pt}
    \renewcommand{\arraystretch}{1.15}
    \caption{FOA encoder pretraining hyperparameters (Stages~1 and~2).}
    \label{tab:app-enc-hparams}
    \begin{tabular}{@{}l c c@{}}
        \toprule
        \textbf{Setting} & \textbf{Stage 1 (semantic)} & \textbf{Stage 2 (localization)} \\
        \midrule
        Training corpus & \multicolumn{2}{c}{$100$k synthetic FOA clips, $1$--$4$ sources, $2$--$8$\,s} \\
        Objective & SigLIP (balanced)~\citep{zhai2023siglip} & $\mathcal{L}_{\text{ang}}+0.5\mathcal{L}_{\text{dist}}+0.5\mathcal{L}_{\text{sem}}$ \\
        Initialization & from scratch & warm-start from Stage 1 \\
        Optimizer & \multicolumn{2}{c}{AdamW ($\beta=0.9,0.95$), weight decay $0.05$, grad clip $1.0$} \\
        Peak / min LR & $3\text{e-}4$ / $1\text{e-}5$ & $1\text{e-}4$ / $1\text{e-}6$ \\
        Schedule / warmup & \multicolumn{2}{c}{cosine, warmup fraction $0.05$} \\
        Batch size (per GPU) & $32$ & $32$ \\
        Epochs / GPUs & $10$ / $2$ & $10$ / $2$ \\
        Precision & \multicolumn{2}{c}{AMP} \\
        \bottomrule
    \end{tabular}
\end{table}

\paragraph{Stage 3: Integration into the Omni backbone.}
We graft the frozen FOA encoder into Qwen3-Omni-30B-A3B (a Mixture-of-Experts model with $\sim\!3$B active parameters). Each FOA clip is processed along two paths: its $W$ channel is down-mixed to mono and fed to the original frozen audio tower to preserve the native semantic audio tokens, while the full four-channel clip is passed through the frozen encoder to produce spatial tokens. A trainable projector (LayerNorm, then $384\rightarrow2048\rightarrow2048$ with GELU) maps the encoder tokens into the language embedding space; the spatial tokens are time-resampled by linear interpolation to match the co-located audio span's token count, and are written into reserved \texttt{<foa\_pad>} placeholder rows bracketed by learnable \texttt{<foa\_bos>}/\texttt{<foa\_eos>} tokens with no vocabulary expansion. Only the projector and the language-model parameters are updated; the FOA encoder, audio tower, and visual tower remain frozen. The final training composition is summarized in Table~\ref{tab:app-data-mix}.

\begin{table}[t]
    \centering
    \fontsize{8pt}{9.2pt}\selectfont
    \setlength{\tabcolsep}{8pt}
    \renewcommand{\arraystretch}{1.15}
    \caption{Stage~3 integration hyperparameters.}
    \label{tab:app-stage3}
    \begin{tabular}{@{}l c@{}}
        \toprule
        \textbf{Setting} & \textbf{Value} \\
        \midrule
        Base model & Qwen3-Omni-30B-A3B~\citep{qwen3omni2025} (MoE, $\sim\!3$B active) \\
        Trainable / frozen & FOA projector $+$ LLM / FOA encoder, audio tower, visual tower \\
        Hardware & $32\times$ A100 ($4$ nodes $\times$ $8$), $\sim\!4$ days \\
        Parallelism & FSDP2, Ulysses seq.-parallel $=4$, expert-parallel $=8$ \\
        Precision & bf16 mixed precision, gradient checkpointing \\
        Global / micro batch size & $128$ / $4$ \\
        Max sequence length & $32{,}768$ \\
        Optimizer & AdamW, weight decay $0.1$, max grad norm $1.0$ \\
        Peak / min LR & $7\text{e-}6$ / $7\text{e-}7$ \\
        Schedule / warmup & cosine, warmup ratio $0.033$ \\
        Epochs & $2$ \\
        \bottomrule
    \end{tabular}
\end{table}

\begin{table}[t]
    \centering
    \fontsize{8pt}{9.2pt}\selectfont
    \setlength{\tabcolsep}{7pt}
    \renewcommand{\arraystretch}{1.15}
    \caption{Training data composition. The navigation examples are sampled from much larger corpus pools with probability $\sim\!0.3$ after action rebalancing.}
    \label{tab:app-data-mix}
    \begin{tabular}{@{}l r r@{}}
        \toprule
        \textbf{Source} & \textbf{Full pool} & \textbf{Used (final mix)} \\
        \midrule
        Spatial audio-visual QA & --- & $26{,}585$ \\
        Spatial audio QA & - & $18{,}518$ \\
        ScaleVLN~\citep{wang2023scalevln} & $403{,}215$ & $139{,}142$ \\
        RxR~\citep{ku2020rxr} & $324{,}342$ & $92{,}418$ \\
        R2R~\citep{anderson2018vision} & $112{,}448$ & $31{,}577$ \\
        VLN-PE~\citep{vlnpe} & $78{,}006$ & $28{,}969$ \\
        In-place rotation data (from ScaleVLN) & --- & $25{,}984$ \\
        
        \midrule
        \textbf{Total} & --- & $363{,}193$ \\
        \bottomrule
    \end{tabular}
\end{table}

\subsection{Evaluation Details}
\label{app:evaluation-details}

\subsubsection{Spatial Audio-Visual QA Evaluation}
\label{app:eval-qa}

Because model outputs are short free-form phrases, we evaluate each prediction with a two-stage protocol that combines lightweight LLM-based extraction with deterministic rule-based judging. In the first stage, the model output is first stripped of any \texttt{<think>} reasoning, and an extractor LLM normalizes the remaining free text into a small set of structured fields: \texttt{yesno} (\texttt{yes}/\texttt{no}), \texttt{dir\_set} (a subset of \{\texttt{left}, \texttt{right}, \texttt{front}, \texttt{back}\}), \texttt{degrees} (float), \texttt{distance\_m} (float, in meters), \texttt{letter} (\texttt{A}/\texttt{B}/\texttt{C}/\texttt{D}), and a free-text \texttt{answer}. Extraction is performed with a text LLM, for which we use \texttt{qwen3.7-flash}. The extractor is queried with a fixed system prompt together with a user message containing only the question and the model answer. The LLM is used solely for answer normalization, not for judging correctness; final correctness is determined entirely in the subsequent rule-based stage, which keeps scoring fully reproducible.

In the second stage, correctness is decided by fixed rules through comparison with the gold reference whenever a task-specific rule applies, as shown in Table~\ref{tab:app-qa-eval}. Multiple-choice questions (\emph{Cognitive Map}) require exact letter match. Directional questions use combinations of the four axis tokens \texttt{front}/\texttt{back}/\texttt{left}/\texttt{right}, giving eight-way resolution; a prediction is correct only if its direction set exactly matches the ground-truth set, with the opposite direction absent. Angular questions require the reported magnitude to fall within a tolerance of $\pm 45^\circ$ and the turn direction to be correct. 3D localization questions require the reported distance to fall within $\pm 2$\,m and the bearing to be correct. Motion questions are scored using the corresponding direction or degree rule depending on subtype. Overall accuracy is computed as the plain average of correctness over all evaluated samples (micro-average), i.e., the total number of correct predictions divided by the total number of samples.

\begin{promptbox}[Extractor system prompt]
You normalize a model's free-text answer for a spatial-audio video QA benchmark. You are given the QUESTION and the MODEL_ANSWER (which may contain reasoning; the real answer is usually its last line). Extract, FROM THE MODEL_ANSWER ONLY, [yesno / dir_set / degrees / distance_m / letter / answer]. Map ahead/forward$\rightarrow$front, behind/rear/backward$\rightarrow$back. For multiple choice, pick the model's FINAL choice, not options it merely discusses. Return ONLY a JSON object.
\end{promptbox}

\begin{table}[h]
    \centering
    \fontsize{8pt}{9.2pt}\selectfont
    \setlength{\tabcolsep}{6pt}
    \renewcommand{\arraystretch}{1.15}
    \caption{QA scoring rules and tolerances by category.}
    \label{tab:app-qa-eval}
    \begin{tabular}{@{}l l@{}}
        \toprule
        \textbf{Category} & \textbf{Correctness rule} \\
        \midrule
        Cognitive Map (A/B/C/D) & exact letter match \\
        Visibility & yes/no matches ground-truth bucket \\
        Source Direction (8-way) & axis set equals GT set, opposite absent \\
        Direction / Camera Rotation (degrees) & $|\Delta\theta|\le45^\circ$ and correct turn direction \\
        3D Localization & $|\Delta \text{dist}|\le2$\,m and correct bearing \\
        Source Motion & direction / degrees rule per subtype \\
        \bottomrule
    \end{tabular}
\end{table}

\subsubsection{Sound-Guided Navigation Evaluation}
\label{app:eval-nav}

\textbf{Closed-loop setup.} Navigation is evaluated in the Habitat simulator using the same episode set (identical start and goal) as the text-instruction baselines. In the sound-guided setting, the agent receives FOA observations together with a short task instruction that identifies the sound event to follow but does not reveal the destination or provide route-level spatial guidance. At each decision step, the agent's live pose selects the nearest receiver in the acoustic field, and the recording's horizontal directional components are remapped into the agent's egocentric frame before being passed (truncated to $10$\,s) to the model server. The model is queried once per decision step and returns $K=8$ relative waypoints $[x_{\text{fwd}},y_{\text{left}},\psi]$; these are converted to discrete actions and executed before the next replan. Evaluation is parallelized by sharding scenes across GPUs, each running an independent model server and worker, and per-shard metrics are merged by an episode-weighted average.

\textbf{Action space and caps.} The discrete action space is \{\texttt{stop}, \texttt{move\_forward}, \texttt{turn\_left}, \texttt{turn\_right}\} with a forward step of $0.25$\,m and a turn angle of $15^\circ$. A waypoint segment is converted to $\text{round}(\Delta\text{pos}/0.25)$ forward steps when its displacement dominates, otherwise to $\text{round}(|\Delta\psi|/15^\circ)$ turn actions. Episodes are capped at $50$ model decisions ($500$ executed steps), and a stuck detector forces termination if net displacement over a $16$-step window stays below $0.25$\,m.

\textbf{Stopping condition.}
All reported navigation experiments use agent self-stop. The model must terminate on its own: a \texttt{stop} is triggered when it emits an empty trajectory or when the maximum waypoint displacement is below $0.3$\,m with negligible yaw (pure in-place turns are executed, not treated as arrival).

\textbf{Metrics.} We report the metrics with a $1$\,m success radius in OmniEchoBench-Nav and $3$\,m success radius in VLN-CE R2R val-unseen, defined in Table~\ref{tab:app-nav-metrics}. NE and TL are averaged with \texttt{nanmean} to exclude errored episodes.

\begin{table}[h]
    \centering
    \fontsize{8pt}{9.2pt}\selectfont
    \setlength{\tabcolsep}{6pt}
    \renewcommand{\arraystretch}{1.15}
    \caption{Navigation metrics The success radius $r_{\mathrm{s}}$ is $1$\,m for OmniEchoBench-Nav and $3$\,m for VLN-CE R2R val-unseen, the same as in prior work.}
    \label{tab:app-nav-metrics}
    \begin{tabular}{@{}l l@{}}
        \toprule
        \textbf{Metric} & \textbf{Definition} \\
        \midrule
        SR $\uparrow$ & success rate: agent stops within $r_{\mathrm{s}}$ of the source \\
        SPL $\uparrow$ & success weighted by (shortest path)/(max(path, shortest path)) \\
        NE $\downarrow$ & navigation error: final distance to goal (m) \\
        OS $\uparrow$ & oracle success: ever within $r_{\mathrm{s}}$ along the trajectory \\
        TL & trajectory length: total path length (m) \\
        \bottomrule
    \end{tabular}
\end{table}

\subsection{Further Experiments}
\subsubsection{Sim2Real Analysis}
\label{app:sim2real}

\begin{table}[h]
    \centering
    \fontsize{8pt}{8.4pt}\selectfont
    \setlength{\tabcolsep}{3.5pt}
    \renewcommand{\arraystretch}{1.15}

    \caption{{\textbf{Sim-to-real transfer results.} OmniEchoBench is partitioned into two scenario-disjoint subsets, with 1,000 samples allocated for training and the remaining samples reserved for testing. Fine-tuning on the real-data subset improves performance on the held-out subset across the evaluated task categories, suggesting that real-data adaptation can help address limitations of training solely on synthetic data.}}

    \label{tab:ablation-sim2real}
    \scalebox{0.8}{
    \begin{tabular}{ c c c c c c c}
        \toprule
        
        \makecell[c]{\textbf{Training with real-world data}} &
        \makecell[c]{\textbf{Camera}\\\textbf{Rotation}} &
        \makecell[c]{\textbf{Source}\\textbf{Direction}} &
        \makecell[c]{\textbf{3D}\\\textbf{Localization}} &
        \makecell[c]{\textbf{Source}\\textbf{Motion}} &
        \makecell[c]{\textbf{Cognitive}\\\textbf{Map}} &
        \makecell[c]{\textbf{Overall}} \\
        \midrule
         \ding{55} & 41.7 & 22.6 & 14.2 & 24.7 & 47.5 & 28.5 \\
        \ding{51} & 44.3 & 31.0 & 21.7 & 35.6 & 48.0 & 34.9 \\
        
        \bottomrule
    \end{tabular}}
\end{table}
Table~\ref{tab:ablation-sim2real} presents the sim-to-real transfer results on OmniEchoBench, where the benchmark is split into training and test subsets. Fine-tuning with real-world data yields a clear improvement in overall performance, raising the average score from 28.5 to 34.9. Performance improves across all task categories, including Camera Rotation (+2.6), Direction (+8.4), 3D Localization (+7.5), Motion (+10.9), and Cognitive Map (+0.5). These results suggest that even a limited amount of real-world data can substantially enhance the model’s spatial-audio reasoning and generalization ability in real environments. More importantly, these results are consistent with a noticeable sim-to-real gap in OmniEchoBench for certain question types, while yielding smaller adaptation gains on other tasks (e.g., Camera Rotation and Cognitive Map).

\subsubsection{VLN Failure Modes} 
\label{app:vln-failure-modes}
\begin{table}[h]
  \centering
  \small
  \caption{OmniEcho on OmniEchoBench-Nav.}
  \label{tab:case3vln}
  \begin{tabular}{llc}
  \toprule
  Aspect & Metric & Value \\
  \midrule
  \multirow{5}{*}{Trajectory}
   & Success Rate (SR)                   & 16.22\% \\
   & SPL                                 & 11.54\% \\
   & Oracle Success (OS, once $\le$1\,m)  & 52.89\% \\
   & Navigation Error (NE)               & 4.09\,m \\
   & Trajectory Length (TL)              & 7.02\,m \\
  \midrule
  \multirow{3}{*}{\shortstack[l]{Direction\\(coarse)}}
   & Step-wise closer rate               & 60.2\% \\
   & Net-approach episode rate           & 64.7\% \\
   & Mean net approach                   & +2.07\,m \\
  \midrule
  \multirow{4}{*}{Distance}
   & MAE $|\text{pred}-\text{true}|$      & 2.20\,m \\
   & Median AE                           & 1.61\,m \\
   & Bias (pred$-$true)                   & +0.70\,m \\
   & Pearson $r$                         & 0.27 \\
  \bottomrule
  \end{tabular}
  \end{table}

  \begin{table}[h]
  \centering
  \small
  \caption{Episode outcome breakdown on OmniEchoBench-Nav (900 episodes).}
  \label{tab:failmode}
  \begin{tabular}{lcc}
  \toprule
  Outcome & Rate & Note \\
  \midrule
  Successful termination within the goal region ($\le$1\,m \& stopped)      & 16.22\% & --- \\
  Entered the goal region without successful termination                 & 36.67\% & OS$-$SR; active-stop failure \\
  Never entered the goal region ($>$1\,m all steps)   & 47.11\% & $100-$OS; navigation failure \\
  \bottomrule
  \end{tabular}
  \end{table}

  \paragraph{Failure-mode analysis.}
  On OmniEchoBench-Nav, the agent moves closer to the source on 60.2\% of decision steps and achieves a net reduction in source distance in 64.7\% of episodes. These statistics indicate a tendency to approach the target, although they do not directly measure sound-direction accuracy. As shown in Table 17, 52.89\% of episodes enter the 1 m goal region at least once, whereas only 16.22\% terminate successfully. Specifically, 36.67\% enter the region without successful termination, and 47.11\% never enter it. Among episodes that enter the goal region, approximately 30.7\% ultimately succeed. This reveals a substantial gap between visiting the target vicinity and completing the task.
  
The outcome breakdown highlights limitations in both target approach and successful termination. Table 16 additionally reports a distance-estimation MAE of 2.20 m and a Pearson correlation of 0.27, indicating limited reliability in estimating target proximity. Under our execution protocol, multiple actions are executed between model decisions, so an agent may enter and leave the goal region before its next opportunity to select a stop. Consequently, the OS–SR gap cannot be attributed exclusively to incorrect stopping decisions. More reliable proximity estimation, adaptive replanning near the target, and a history of recent acoustic observations are promising directions for improving task completion; their individual effects remain to be evaluated.

\subsubsection{VLN with GT direction}
\label{app:vln-gt-direction}

\begin{table}[h]
    \centering
    \fontsize{7pt}{8.4pt}\selectfont
    \setlength{\tabcolsep}{3.5pt}
    \renewcommand{\arraystretch}{1.15}
    \caption{\textbf{Performance of OmniEcho with GT direction on our OmniEchoBench-Nav.} OmniEcho$^{\dagger}$ receives the GT egocentric source direction at each decision step under the same action-execution and agent self-stopping protocol. The GT-direction variant achieves higher SR and SPL but lower OS, showing different effects on successful task completion and goal-region visitation.}
    \label{tab:gt-direction}

    \begin{tabular}{@{}l c c c c c c c@{}}
        \toprule
        \multirow{2}{*}{\textbf{Model}}
        & \multirow{2}{*}{\textbf{Base Model}}
        & \multirow{2}{*}{\textbf{Guidance}}
        & \multicolumn{5}{c}{\textbf{Val-Unseen}}
        \\
        \cmidrule(lr){4-8}

        & & & \textbf{SR$\uparrow$}
        & \textbf{SPL$\uparrow$}
        & \textbf{NE$\downarrow$}
        & \textbf{OS$\uparrow$}
        & \textbf{TL}
        \\

        \midrule
        OmniEcho
        & Qwen3-Omni-30B-A3B
        & FOA audio
        & {16.2}
        & {11.5}
        & 4.09
        & 52.89
        & 7.02
        \\

        OmniEcho$^{\dagger}$
        & Qwen3-Omni-30B-A3B
        & GT direction
        & {23.7}
        & {22.1}
        & 4.14
        & 34.33
        & 3.47
        \\

        \bottomrule
    \end{tabular}
\end{table}

To examine how ground-truth directional guidance affects closed-loop navigation, we evaluate a GT-direction variant of OmniEcho that receives the target’s ground-truth egocentric direction at each decision step. The action-execution and agent self-stopping protocols remain unchanged. As shown in Table 18, this variant increases SR from 16.2\% to 23.7\% and SPL from 11.5\% to 22.1\%. However, OS decreases from 52.89\% to 34.33\%, while NE remains similar (4.09 m versus 4.14 m).
These results are compatible because OS measures whether a trajectory ever enters the goal region, whereas SR additionally requires successful termination there. Among episodes that enter the region, the fraction ending successfully increases from approximately 30.7\% to 69.0\%. This is a descriptive comparison over the trajectories generated by each variant; it does not isolate stopping performance on a common set of reached episodes. The reduction in mean TL from 7.02 m to 3.47 m may reflect more direct successful trajectories, earlier termination of unsuccessful episodes, or both. Thus, the benefits of ground-truth directional guidance are primarily reflected in successful termination and path efficiency, rather than an increased likelihood of entering the goal region.


\subsection{Visualization Examples}

\begin{figure*}[h!]
    \centering
    \includegraphics[width=\textwidth]
    {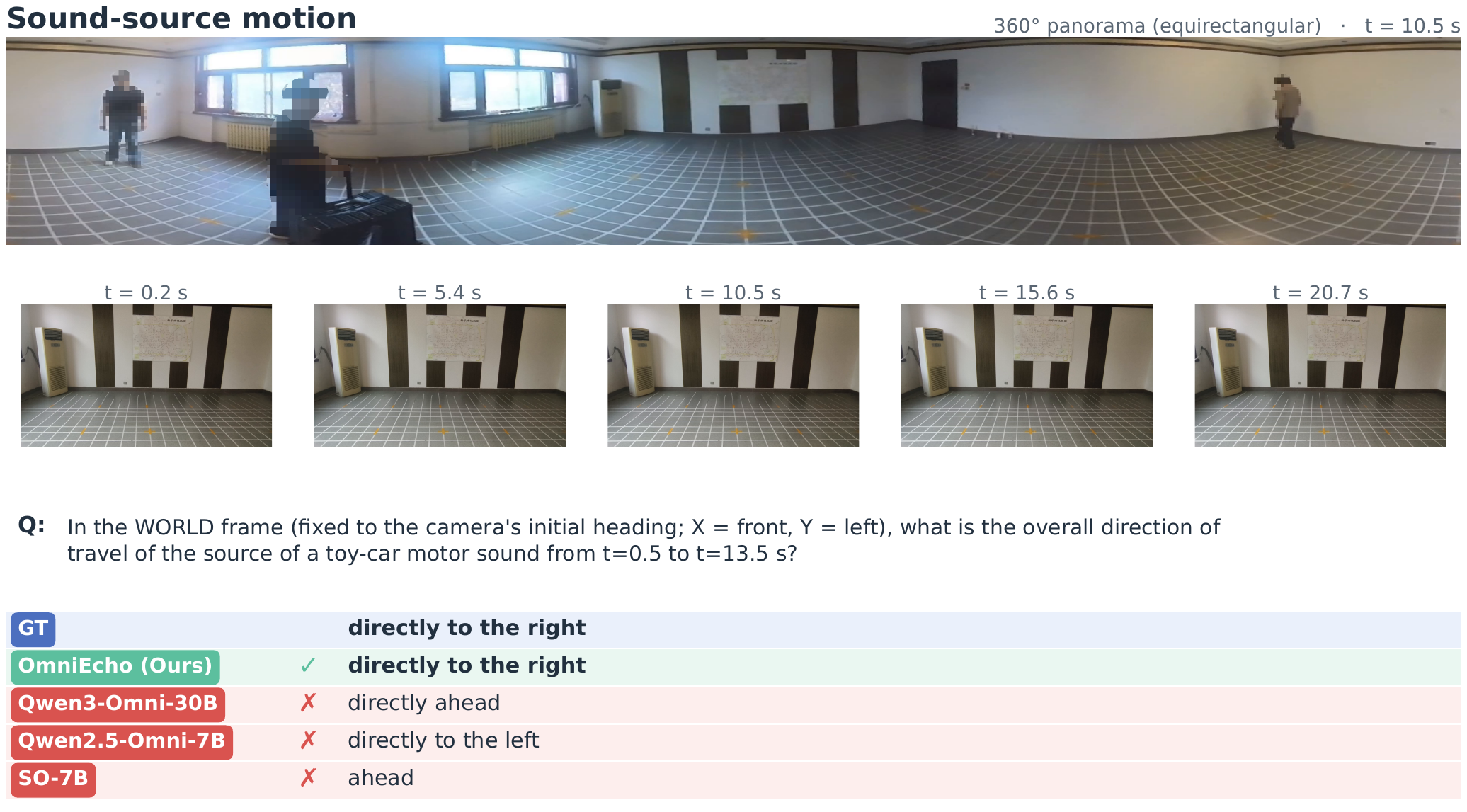}
    \caption{A successful sound-source motion example from OmniEchoBench-QA.}
    \label{fig:qa-success-case-1}
\end{figure*}

\begin{figure*}[h]
    \centering
    \includegraphics[width=\textwidth]
    {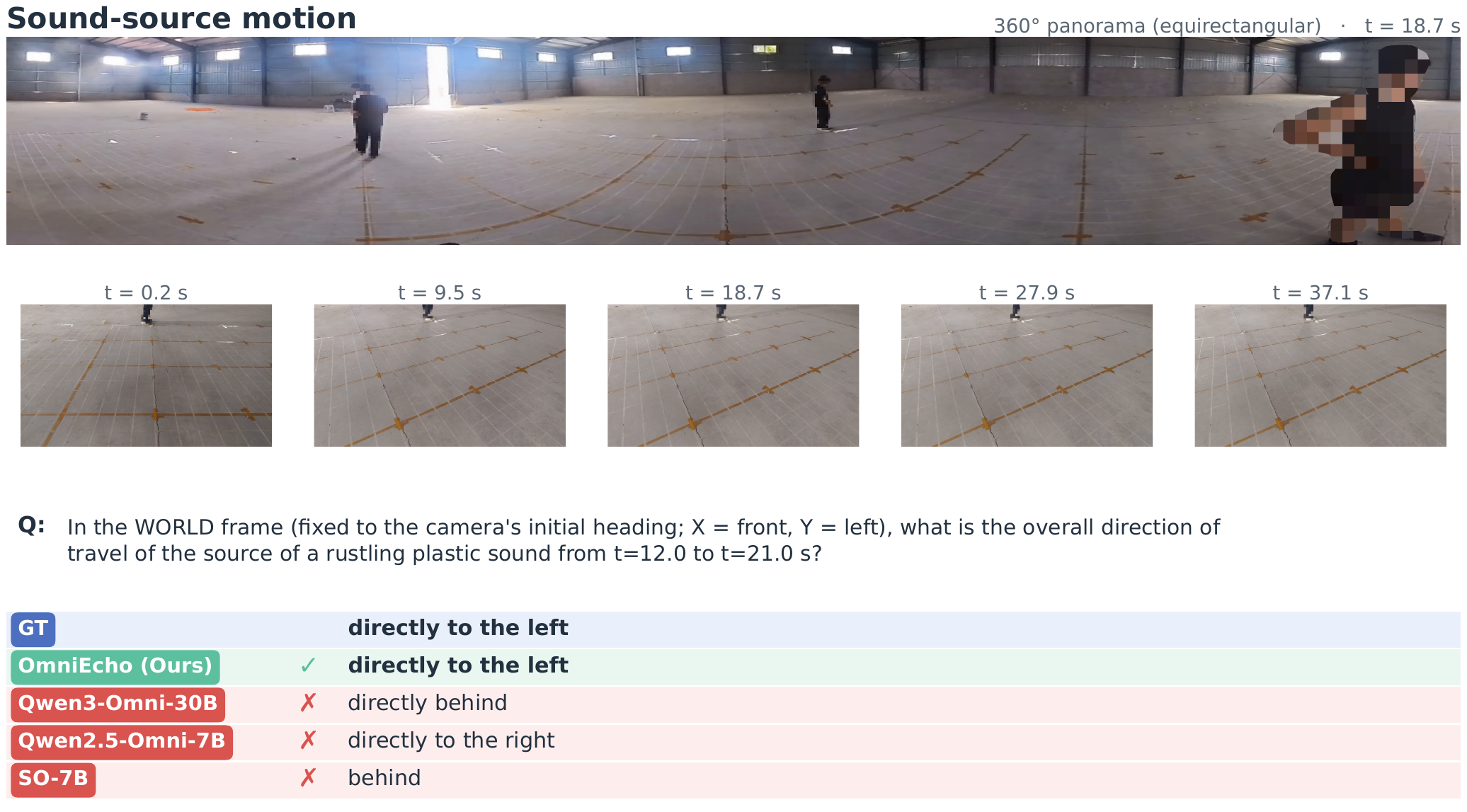}
    \caption{A successful sound-source motion example from OmniEchoBench-QA.}
    \label{fig:qa-success-case-2}
\end{figure*}

\begin{figure*}[h]
    \centering
    \includegraphics[width=\textwidth]
    {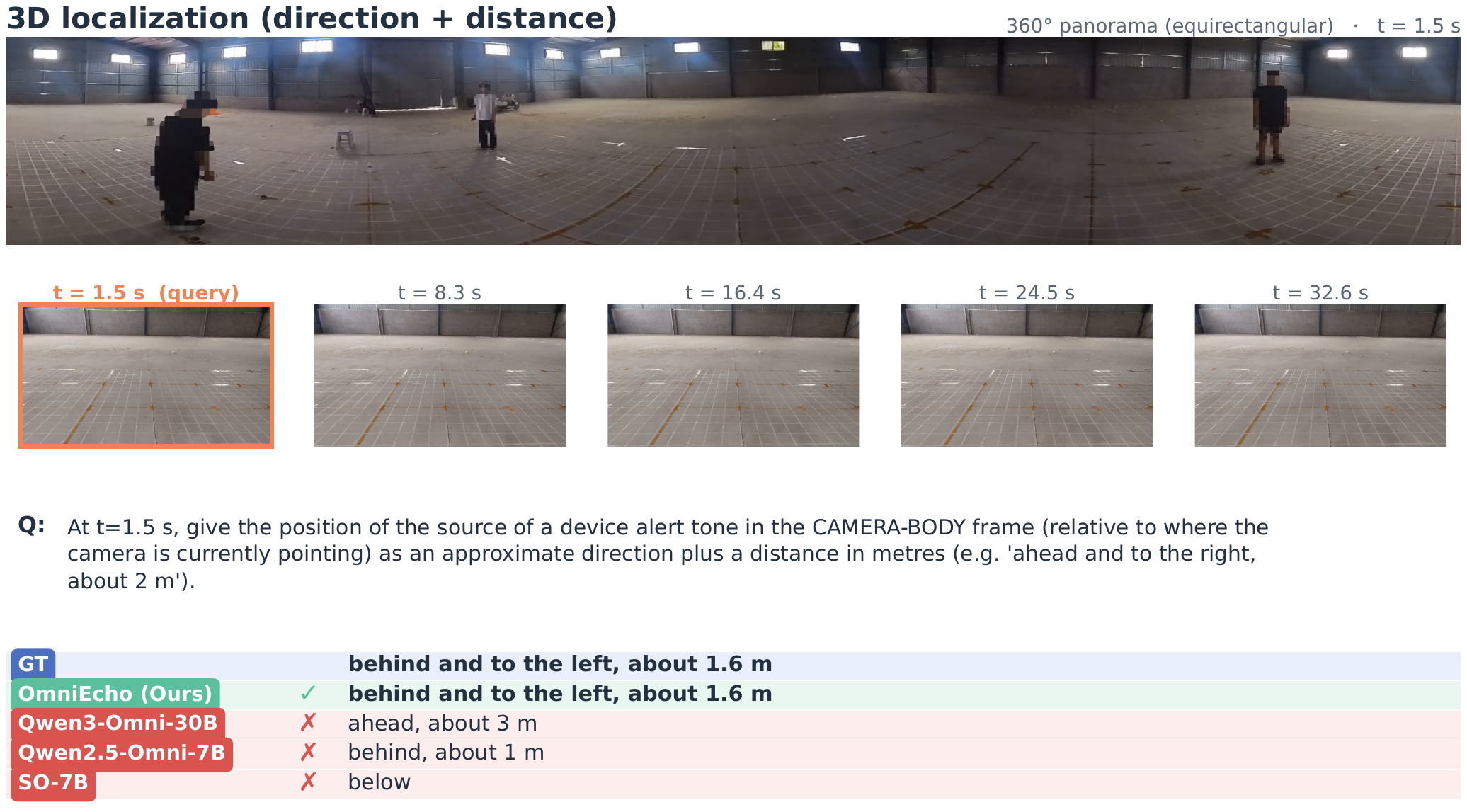}
    \caption{A successful 3D localization example from OmniEchoBench-QA.}
    \label{fig:qa-success-case-3}
\end{figure*}

\begin{figure*}[h]
    \centering
    \includegraphics[width=\textwidth]
    {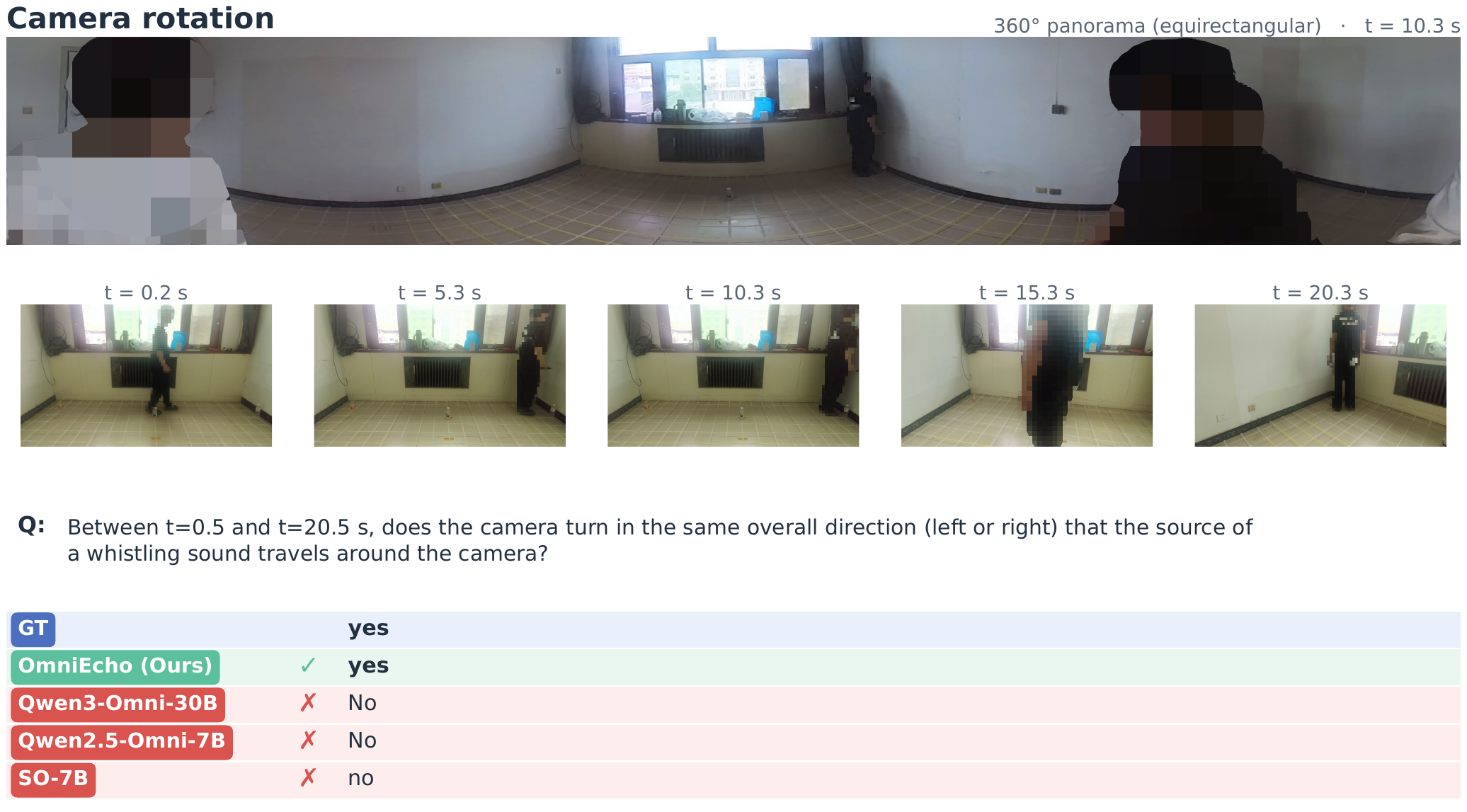}
    \caption{A successful camera-rotation example from OmniEchoBench-QA.}
    \label{fig:qa-success-case-4}
\end{figure*}

\begin{figure*}[h]
    \centering
    \includegraphics[width=\textwidth]
    {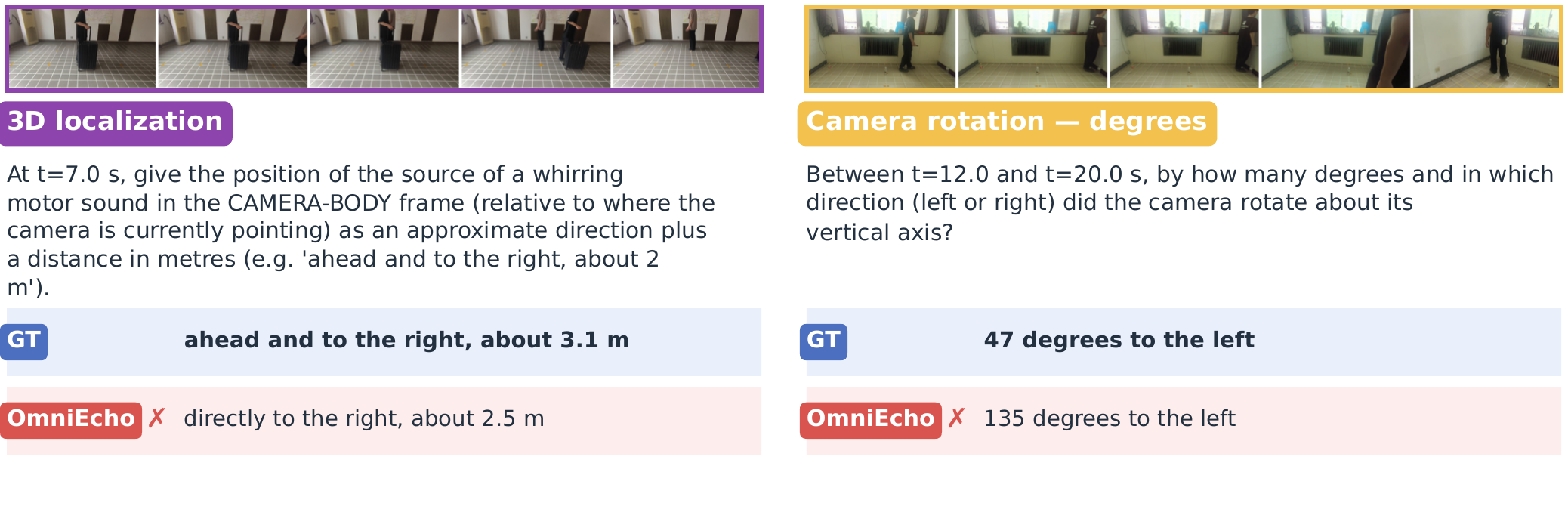}

    \caption{Selected failure cases on OmniEchoBench-QA.
    OmniEcho preserves part of the correct coarse orientation but misses finer
    geometric details: it predicts an ahead-right source as directly to the right
    and estimates a $135^\circ$ leftward camera rotation instead of $47^\circ$.}
    \label{fig:qa-failure-cases}
\end{figure*}

Figures~\ref{fig:qa-success-case-1}--\ref{fig:qa-success-case-4} visualize representative predictions on OmniEchoBench-QA. The successful examples cover sound-source motion, 3D localization, and camera rotation, illustrating the model's ability to recover spatial states from FOA audio and visual observations. Figure~\ref{fig:qa-failure-cases} shows two failure examples. In the 3D localization example, OmniEcho captures the coarse direction but predicts the source as directly to the right rather than ahead-right and underestimates its distance. In the camera-rotation example, it identifies the correct rotation direction but substantially overestimates the angle. Together, these visualizations highlight the remaining gap between coarse spatial understanding and fine-grained geometric estimation.

\begin{figure*}[h]
    \centering
    \includegraphics[width=\textwidth]
    {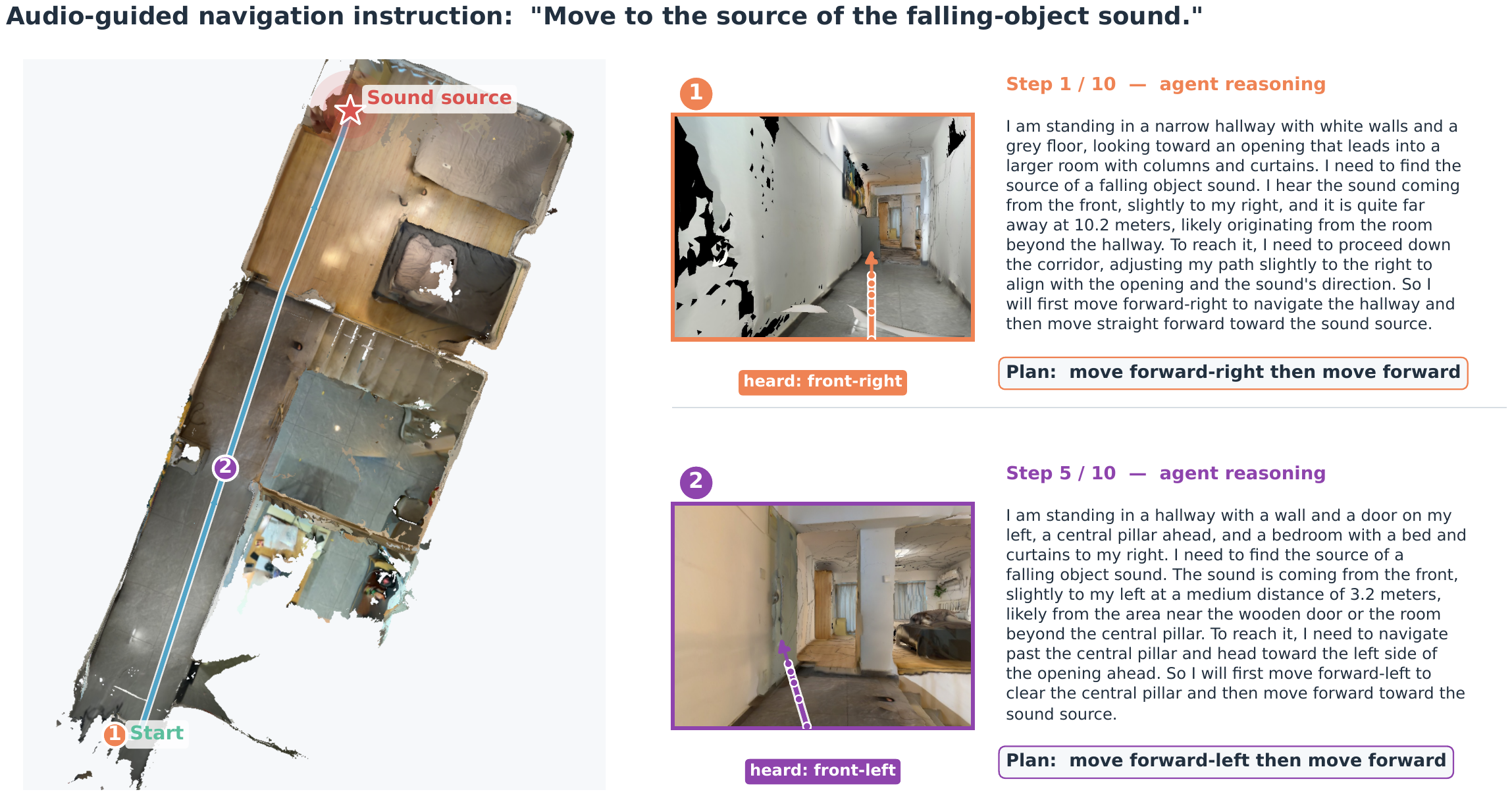}
    \caption{Selected successful navigation trajectories on OmniEchoBench-Nav.}
    \label{fig:nav-success-case-1}
\end{figure*}

\begin{figure*}[h]
    \centering
    \includegraphics[width=\textwidth]
    {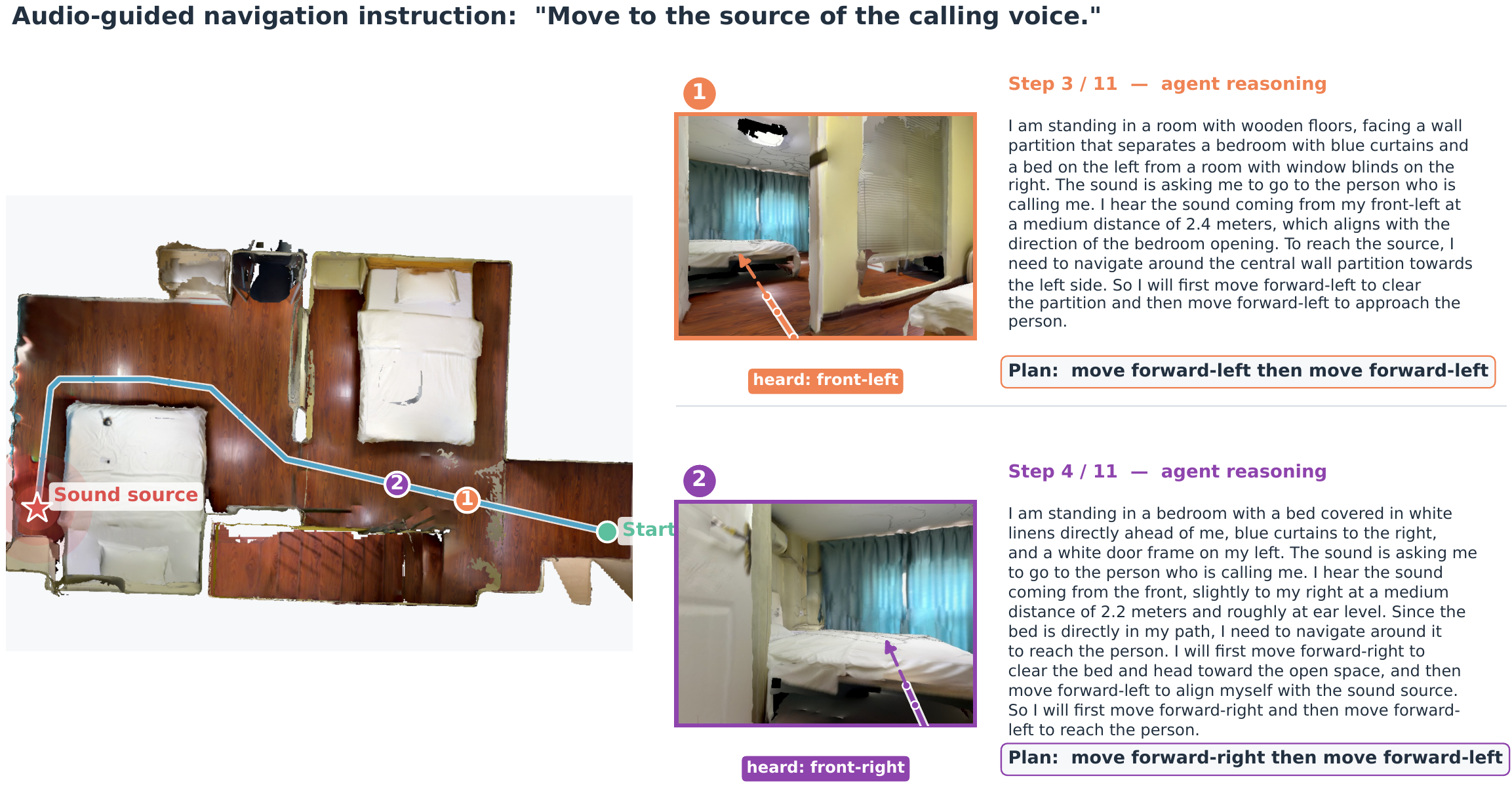}
    \caption{Selected successful navigation trajectories on OmniEchoBench-Nav.}
    \label{fig:nav-success-case-2}
\end{figure*}

\begin{figure*}[h]
    \centering
    \includegraphics[width=\textwidth]
    {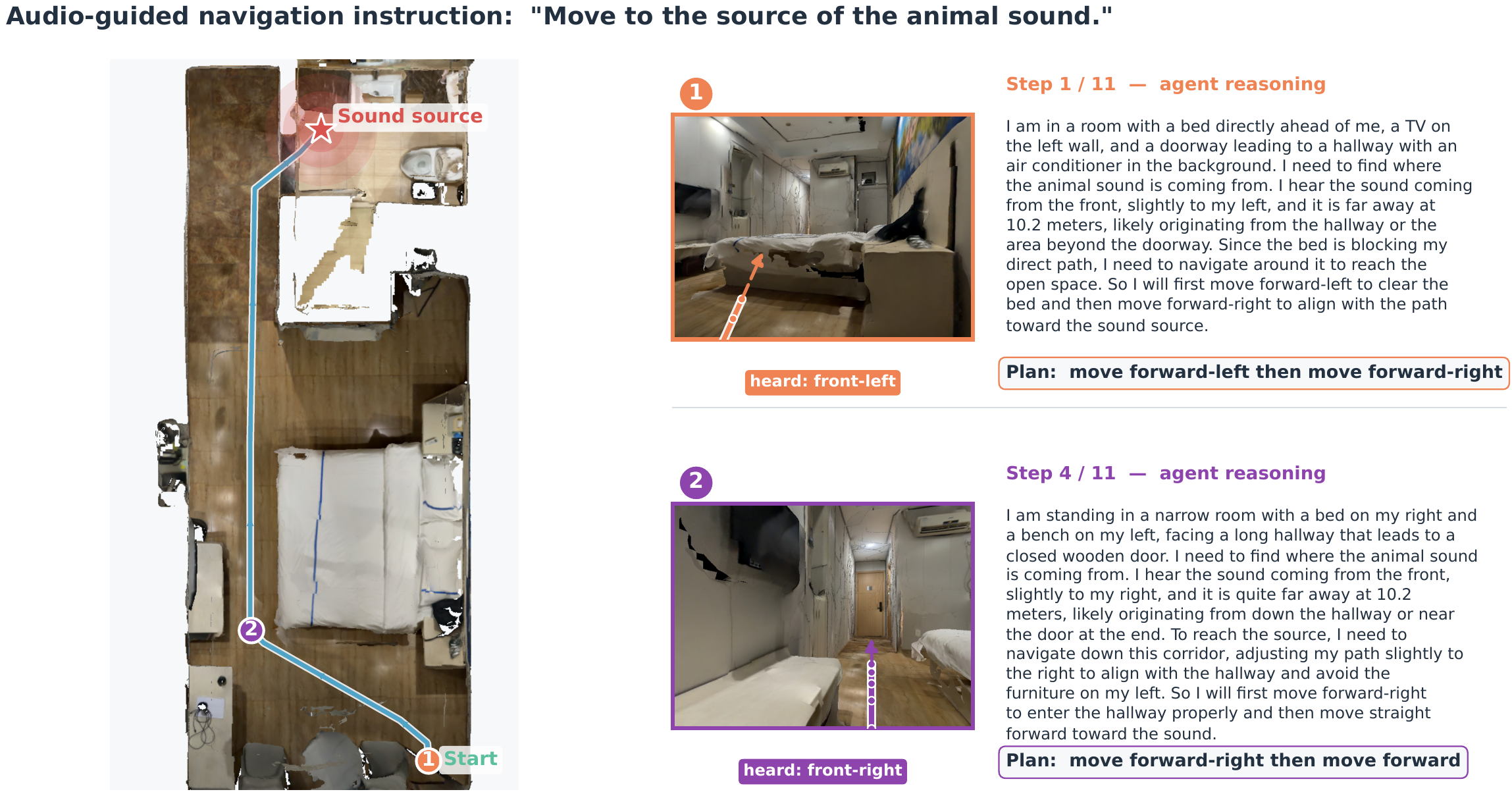}
    \caption{Selected successful navigation trajectories on OmniEchoBench-Nav.}
    \label{fig:nav-success-case-3}
\end{figure*}

\begin{figure*}[h]
    \centering
    \includegraphics[width=\textwidth]
    {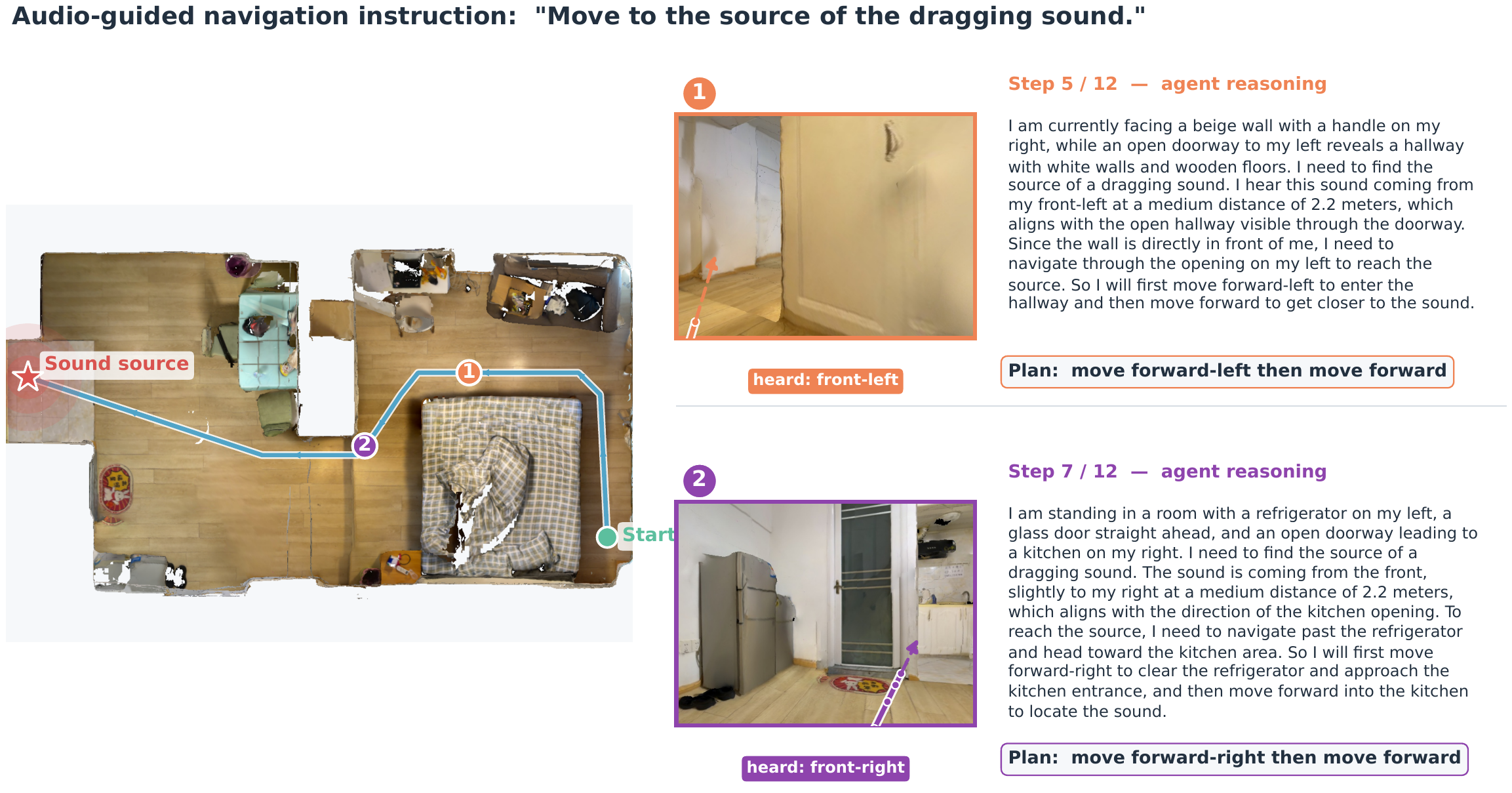}
    \caption{Selected successful navigation trajectories on OmniEchoBench-Nav.}
    \label{fig:nav-success-case-4}
\end{figure*}

\begin{figure*}[h]
    \centering
    \includegraphics[width=\textwidth]{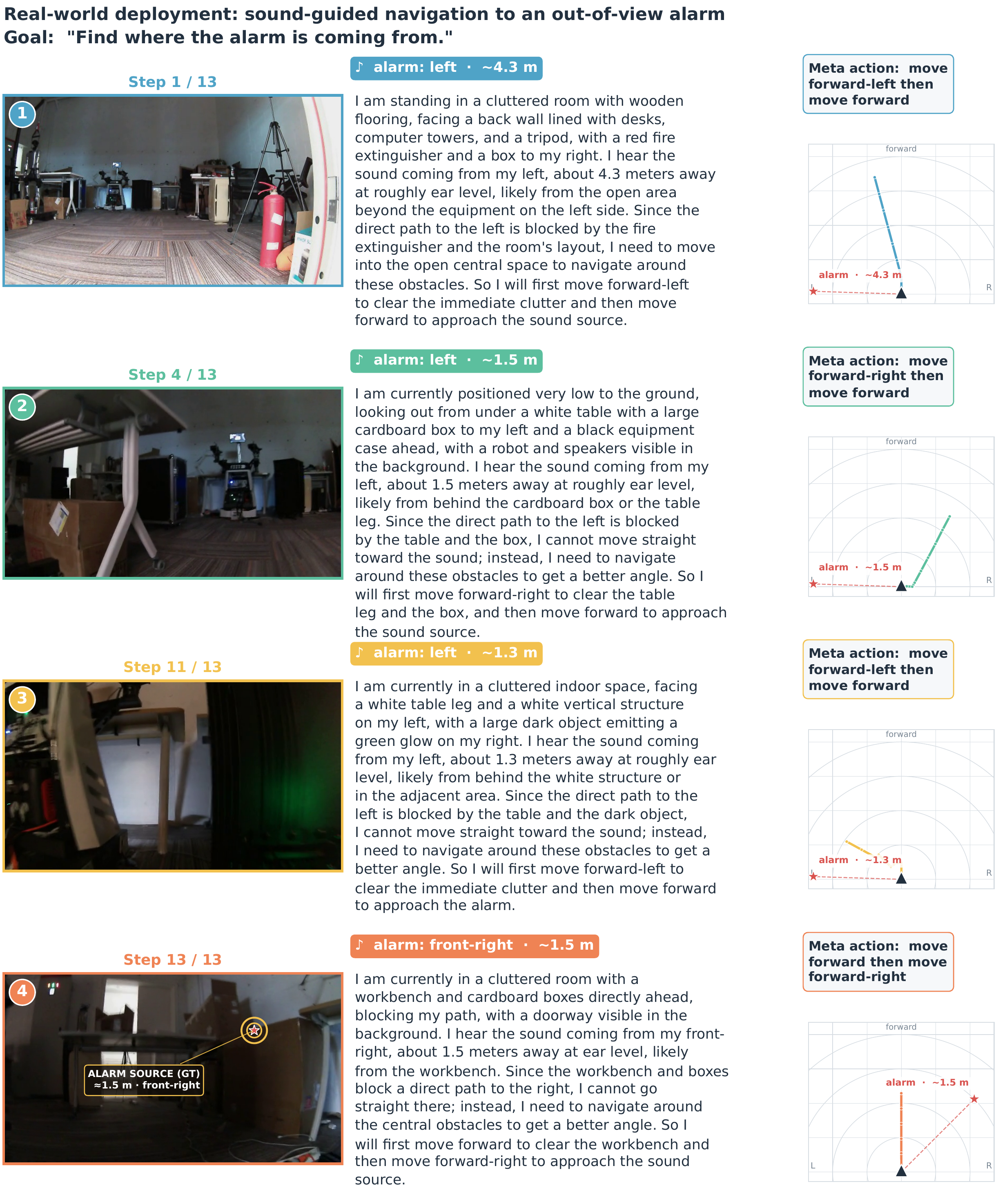}
    \caption{Qualitative example of OmniEcho deployed on a physical robot for sound-guided navigation. The robot progressively localizes and approaches an initially out-of-view alarm by jointly reasoning over egocentric spatial audio and visual observations while navigating around obstacles.}
    \label{fig:nav-demo}
\end{figure*}

Figures~\ref{fig:nav-success-case-1}--\ref{fig:nav-success-case-4} visualize successful closed-loop episodes from four OmniEchoBench-Nav scenes. Each example pairs the global trajectory with intermediate egocentric observations, estimated sound directions, and planned actions, showing how spatial-audio cues are converted into scene-conditioned movement. Figure~\ref{fig:nav-demo} provides a further visualization on a physical robot. Across successive steps, the robot updates the estimated direction and distance of an initially out-of-view alarm, plans around intervening obstacles using spatial audio and visual observations, and progressively approaches the source.

\end{document}